\documentclass[pra,twocolumn,superscriptaddress,showpacs,preprintnumbers,amsmath,amssymb,floatfix]{revtex4-2}

\usepackage{graphicx}
\usepackage{dcolumn}
\usepackage{bm}
\usepackage{enumerate}
\usepackage{mathdots}
\usepackage{color}

\begin{document}
\title{Tunable quantum interference effects in Floquet two- and three-level systems}

\author{Yingying Han}
\affiliation{Shenzhen Institute for Quantum Science and Engineering, Southern University of Science and Technology, Shenzhen 518055, China}
\affiliation{Shenzhen Key Laboratory of Ultraintense Laser and Advanced Material Technology, Center for Advanced Material Diagnostic Technology, and College of Engineering Physics, Shenzhen Technology University, Shenzhen 518118, China}

\author{Minchen Qiao}
\affiliation{Key Laboratory of Artificial Micro- and Nano-structures of Ministry of Education, and School of Physics and Technology, Wuhan University, Wuhan, Hubei 430072, China}
\affiliation{School of integrated circuits, Tsinghua University, Beijing 100084, China}

\author{Xiao-Qing Luo}
\affiliation{Hunan Province Key Laboratory for Ultra-Fast Micro/Nano Technology and Advanced Laser Manufacture, School of Electrical Engineering, University of South China, Hengyang, 421001, China;}

\author{Tie-Fu Li}
\affiliation{School of Integrated Circuits and Frontier Science Center for Quantum Information, Tsinghua University, Beijing 100084, China}
\affiliation{Beijing Academy of Quantum Information Sciences, Beijing 100193, China}

\author{Wenxian Zhang}
\affiliation{Key Laboratory of Artificial Micro- and Nano-structures of Ministry of Education, and School of Physics and Technology, Wuhan University, Wuhan, Hubei 430072, China}

\author{Xiu-Hao Deng}
\email[Corresponding email: ]{dengxh@sustech.edu.cn}
\affiliation{Shenzhen Institute for Quantum Science and Engineering, Southern University of Science and Technology, Shenzhen 518055, China}
\affiliation{Guangdong Provincial Key Laboratory of Quantum Science and Engineering,
Southern University of Science and Technology, Shenzhen, 518055, China}

\author{J.Q. You}
\affiliation{Interdisciplinary Center of Quantum Information, State Key Laboratory of Modern Optical Instrumentation, and Zhejiang Province Key Laboratory of Quantum Technology and Device, Department of Physics, Zhejiang University, Hangzhou 310027, China}

\author{Dapeng Yu}
\affiliation{Shenzhen Institute for Quantum Science and Engineering, Southern University of Science and Technology, Shenzhen 518055, China}
\affiliation{Guangdong Provincial Key Laboratory of Quantum Science and Engineering,
Southern University of Science and Technology, Shenzhen, 518055, China}

\date{\today}

\begin{abstract}
Quantum interference effects in the un-modulated quantum systems with light-matter interaction have been widely studied, such as electromagnetically induced transparency (EIT) and Autler-Townes splitting (ATS). However, the similar quantum interference effects in the Floquet systems (i.e., periodically modulated systems), which might cover rich new physics, were rarely studied. In this article, we investigate the quantum interference effects in the Floquet two- and three-level systems analytically and numerically. We show a coherent destruction tunneling effect in a lotus-like multi-peak spectrum with a Floquet two-level system, where the intensity of the probe field is periodically modulated with a square-wave sequence. We demonstrate that the multi-peak split into multiple transparency windows with tunable quantum interference if the Floquet system is asynchronously controlled via a third level. Based on phenomenological analysis with Akaike information criterion, we show that the symmetric central transparency window has a similar mechanism to the traditional ATS or EIT depending on the choice of parameters, additional with an extra degree of freedom to control the quantum interference provided by the modulation period. The other transparent windows are shown to be asymmetric, different from the traditional ATS/EIT windows. These non-trivial quantum interference effects open up a new scope to explore the applications of the Floquet systems.

\end{abstract}

\maketitle

\section{Introduction}
Floquet systems, which could be characterized by periodically modulated Hamiltonian, display rich dynamics and novel phenomena that are absent in their un-modulated counterparts, such as quantum dynamical decoupling~\cite{PhysRevLett.114.190502, PhysRevB.75.201302, PhysRevB.77.125336}, time crystal~\cite{PhysRevLett.117.090402, SachaTime}, Mach-Zehnder
interferometer~\cite{Oliver2005Mach,Shevchenko20101}, time-domain Fresnel lenses~\cite{PhysRevLett.89.203003}, time-domain grating~\cite{HanPRApplied} and Floquet topological phase~\cite{PhysRevB.102.041119}. The studies of periodically modulated systems are also known as Floquet engineering ~\cite{2014Universal,PhysRevApplied.14.054049,PhysRevA.77.053601,PhysRevA.102.012221,PhysRevA.101.022108}. Moreover, for the Floquet systems, the dynamic steady-states are periodic steady-states, which emerge in a balance of the energy injection by the periodic driving and the relaxation processes~\cite{PhysRevA.93.013820,2020General,Han:22}. To explore the stability properties of the periodic steady-states,  the time-average values of observable physical quantities are usually observed experimentally~\cite{PhysRevA.105.012418,HanPRApplied}, and stroboscopic evolution of a periodically driven quantum system in steps of the modulation period is usually adopted in theoretical studies~\cite{PhysRevA.93.032121}.

Quantum interference effect (QIE) is the key to the quantum nature of a system. In an atom-field interacting quantum system, QIE enables rich interesting physics and applicable phenomena, such as electromagnetically induced transparency (EIT)~\cite{PhysRevLett.64.1107,PhysRevA.81.063823} and Autler-Townes splitting (ATS)~\cite{PhysRev.100.703}. Both of them are observed with a transparency window induced by the un-modulated coherent drive fields, though they originate from controversial mechanisms that have been studied over decades in various systems and scenarios ~\cite{2011Observation,PhysRevA.81.053836,PhysRevA.87.043813}. In the above un-modulated systems, the properties of QIE can be adjusted by manipulating the properties of the system, which is relatively difficult. Few studies have been done to alter the properties of QIE by using tunable auxiliary energy levels or phase-modulated fields~\cite{PhysRevA.89.063822,2022Phase,PhysRevLett.96.183601}. Recently, with the rising interest in Floquet systems, rich novel physics are discovered, where the Floquet parameters (such as modulation period, modulation scheme, etc.) may be useful tools for tuning the properties of QIEs. However, to our knowledge, the QIEs in the Floquet controlled systems have not been adequately studied~\cite{PhysRevA.49.1950}, especially the effect of the modulation period on the properties of QIEs.

In this work, we first study the Floquet two-level system, where the intensity of the probe field has the form of square-wave periodic sequence, which is a basic model of direct frequency comb spectroscopy~\cite{Picqu2019Frequency} and is different from the sine/cosine pulse trains in the Mach-Zehnder interferometer~\cite{Oliver2005Mach,PhysRevLett.112.027001}. Note that here the modulation period is shorter than the system's coherence time. We explore the steady excitation probability over a wide driving strength range and give the analytically results with some special scenarios. A lotus-like multi-peak spectrum is observed and the coherent destruction of tunneling effect is found in the strong driving regime. When the Floquet two-level system assisted with a third energy level and a periodically modulated control field, each peak splits into two, resulting in multiple transparency windows. Here the modulation pulse of the control field is the same as that of the probe field, but they are asynchronous. An intriguing finding is that the central transparency window (CTW) in the Floquet three-level system has a similar profile to the traditional EIT or ATS in the un-modulated systems. We use the Akaike information criterion (AIC) method~\cite{PhysRevLett.107.163604} to discern the CTW from EIT and ATS by evaluating their relative AIC weights for different modulation periods and find that the CTW could be EIT-like or ATS-like in different parameter regimes. Moreover, the quantum interference of the CTW can be modified by the modulation period, which as an additional degree of freedom increases the tunable space of CTW. Therefore, the CTW may provide a superior platform for the explosion of quantum technology than the traditional EIT/ATS.

This work is outlined as follows. We explore the QIEs in the Floquet two-level system in Sec.~\ref{sec:two} and Floquet three-level system in Sec.~\ref{sec:three}. Conclusions and discussions are given in the last section of the article.

\section{The Floquet two-level system}\label{sec:two}
\begin{figure}[pth]
\centering
\includegraphics[width=3.2in]{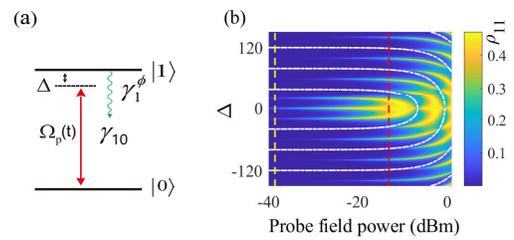}
\caption{(a) Schematic of a two-level system. A probe-field couples levels $|0\rangle$ and $|1\rangle$ with a periodically modulated Rabi frequency $\Omega_p(t)=\Omega_p(t+\tau)$ and a detuning $\Delta$. $\gamma_{10}$ is the population damping rate from level $|1\rangle$ to $|0\rangle$. $\protect\gamma_{1}^{\protect\phi}$ is the dephasing rate of state $|1\rangle$. (b) The steady excitation probability $\protect\rho_{11}$ as a function of detuning and the probe field power. The graph is obtained by numerically calculating the Lindblad master equation in Eq.~(\ref{eq:lmetwo}) for $\protect\gamma _{10}=1$, $\protect\gamma_{1}^{\protect\phi}=0.4$, and $\protect\tau=0.05$. Note that in this work we normalized the parameters in terms of $\gamma_{10}$, such as $\protect\gamma_{1}^{\protect\phi}/\gamma _{10}=0.4$, $\protect\tau~\gamma_{10}=0.05$, and they are simplified to $\protect\gamma_{1}^{\protect\phi}=0.4$ and $\protect\tau=0.05$. The white dashed lines show the excitation minima [i.e.,$\protect\sqrt{\Omega_{p}^{2}+\Delta^{2}}=2n\protect\omega,~(n=1,2\cdots)$] with modulation frequency $\omega=1/\tau$. The vertical red dashed line at $\Omega_{p}=\omega=-13\mathrm{dBm}$ divides the weak and strong coupling ranges. The vertical yellow dashed line denotes $\Omega_{p}=1\approx-38.6\mathrm{dBm}$.}
\label{fig:modtwo}
\end{figure}

In this section, we first explore the QIEs in a Floquet two-level system denoted by $|0\rangle$ and $|1\rangle $ with energy $\omega _{0,1}$, as shown in Fig.~\ref{fig:modtwo}(a). A probe field with frequency $\omega _{p}$ and periodically modulated Rabi frequency $\Omega_p(t)=\Omega_p(t+\tau)$ couples levels $|0\rangle$ and $|1\rangle$. Here $\tau$ is the modulation period and the detuning between the transition frequency (i.e., $\omega _{10}=\omega _{1}-\omega _{0}$) and the frequency of the probe field is $\Delta=\omega_{10}-\omega_p$. To simplify the calculation, we consider the square-wave periodic sequence as the modulation scheme in this work, i.e.,
\begin{eqnarray}  \label{eq:omep}
\Omega_{p}(t)=\left\{
\begin{array}{ll}
\Omega_p, & ~~~~~t\in[n\tau,(n+\frac{1}{2})\tau] \\
0, & ~~~~~t\in[(n+\frac{1}{2})\tau,(n+1)\tau]%
\end{array}
\right.
\end{eqnarray}
with $n=0,1,2\cdots$. The Hamiltonian of such system is
\begin{eqnarray}\label{eq:two}
  H=\Delta (-|0\rangle \langle 0|+|1\rangle\langle 1|)/2-[\Omega _{p}(t)|0\rangle \langle1|+h.c.]/2,
\end{eqnarray}
with $\hbar=1$. Note that for our time-dependent Hamiltonian, here we apply rotating-wave approximation (RWA) by assuming $\omega=1/\tau\ll 2\omega_{10}$, $\Omega_p\ll\omega_{10}$, which are also the parameter ranges of many experimental studies. Our earlier experimental work has shown
the validity of the RWA with a periodically driven superconducting qutrit~\cite{HanPRApplied}. Moreover, in Appendix~\ref{app:rwa}, we also theoretically verified the validity of RWA in detail. With the assumption of Markovian noise background, the Floquet systems' density matrix evolves as the Lindblad master equation~\cite{PhysRevB.104.165414,PhysRevB.101.100301},
\begin{eqnarray}\label{eq:lmetwo}
\dot{\rho}&=&-i[H(t),\rho ]+\frac{\gamma _{10}}{2}(2\sigma
_{01}\rho \sigma _{10}-\sigma _{11}\rho -\rho \sigma _{11})\nonumber\\
&+&\gamma _{1}^{\phi }(2\sigma _{11}\rho \sigma _{11}-\sigma
_{11}\rho -\rho \sigma _{11}),
\end{eqnarray}%
where $\sigma _{ij}=|i\rangle \langle j|$ ($i,j=0,1$) is the projection operators and $H(t)$ is shown in Eq.~(\ref{eq:two}). To observe the steady-state characteristics of the periodically driven systems, we only observe the data at the end of each modulation period (i.e., the time evolution step is $\tau$), and ignore the micro-dynamics within one modulation period. The dynamics start from the ground state $|0\rangle$ and then evolve to the steady states (see also Fig.~\ref{fig:valid1}), which are observed to study the steady-state characteristics of the periodically driven systems.

Figure.~\ref{fig:modtwo}(b) shows the steady excitation probability $\protect\rho_{11}$ as a function of detuning and the probe field power. Note that here the unit of the probe field is $\mathrm{dBm}$, where $1\mathrm{dBm}=10\mathrm{log(C}~\Omega _{p}^{2})$ with $\mathrm{C}=1.38\times 10^{-4}$ determined by experimental data~\cite{HanPRApplied, PhysRevA.93.053838}. We find multi-peak phenomenon in the weak driving range ($\Omega_{p}<-13\mathrm{dBm}$) and the sidebands are well separated, which is caused by the Fourier components of the square-wave modulated $\Omega_p(t)$ in Eq.~(\ref{eq:omep}), i.e.,
\begin{equation}\label{eq:foueop}
\Omega _{p}(t)=\frac{\Omega _{p}}{2}-\sum_{n=1}^{\infty }(-1)^{n}\Omega_{pn}\cos (\omega _{n}t),
\end{equation}
where
\begin{eqnarray}
\Omega _{pn}&=&\frac{\Omega _{p}}{(2n-1)\pi},\\
\omega _{n}&=&(2n-1)\omega,(n=1,2,3\cdots ).
\end{eqnarray}
These expressions clearly demonstrate that the square-wave modulated field is equivalent to employing many frequency-tunable fields and the distance between them can be adjusted by the modulation frequency $\omega$ with frequency separation $\omega_{n+1}-\omega_{n}=2\omega$.

In the strong driving range ($\Omega_{p}>-13\mathrm{dBm}$), the peaks get broader and overlap with each other, eventually forming a lotus pattern, which is too complex to get the analytic solution. However, we can obtain the steady solutions of $\rho _{11}$ with $\Delta =0$ by
calculating the optical Bloch equations (see Appendix~\ref{app:twor}),
\begin{align}
\rho _{11} =\frac{1}{2}-\frac{\gamma _{10}}{2}\sum_{n=-\infty }^{\infty }
\frac{\gamma _{1}\Omega _{n}^{2}}{\gamma _{1}^{2}+(\Omega _{p}/2-n\omega
)^{2}},  \label{eq:two11}
\end{align}
where $\gamma _{1}=3\gamma _{10}/4+\gamma _{1}^{\phi}/2 $ and n are integers. $\Omega _{n}$ is a series of Bessel functions with variable $\Omega_{p}/\omega $ [see Eq.~(\ref{eq:omega})]. When $\tau \rightarrow 0$, $\Omega _{0}\rightarrow 1$ and $\Omega _{n\neq 0}\rightarrow 0$, then Eq.~(\ref{eq:two11}) can be reduced to $\rho _{11}\approx \frac{1}{2}-\frac{\gamma _{10}}{2}
\frac{\gamma _{1}}{\gamma _{1}^{2}+(\Omega _{p}/2)^{2}}$, which is close to the conventional result of the un-modulated system $\rho _{11}=\frac{1}{2}-\frac{\gamma _{10}}{2}\frac{\gamma _{1}'}{\gamma _{1}'\gamma _{10}+\Omega _{p}^{2}}$ with $\gamma _{1}^{\prime }=\gamma _{10}/2+\gamma_{1}^{\phi }$, except the effective Rabi frequency of the probe field to be $\Omega _{p}/2$. These indicate that when $\tau \rightarrow 0$, the central peak of the multi-peak phenomenon is similar to the single resonant peak in the un-modulated system. However, when $\tau $ away from $0$, $\Omega _{n}$ significantly modifies the signal and induces a non-trivial phenomenon, coherent destruction of tunneling~\cite{PhysRevLett.67.516, PhysRevA.75.063414,PhysRevA.91.063804}, where the steady excitation probability $\rho_{11}$ is partially suppressed and arising from the
superposition of degenerate Floquet states. When $\Delta=0$, from Eq.~(\ref{eq:two11}) one finds that the conditions for $\rho_{11}$ to take the local minimum values are $\Omega_p=2n\omega ,~(n=1,2\cdots )$. Expanding to the more general cases, the positions of the coherent destruction of tunneling are determined by the relationship between the effective Rabi frequency and the modulation frequency, i.e., $\sqrt{\Omega _{p}^{2}+\Delta ^{2}}=2n\omega ,~(n=1,2\cdots )$, as the white dashed lines shown in Fig.~\ref{fig:modtwo}(b).

\section{The Floquet three-level system}\label{sec:three}
\begin{figure}[pth]
\includegraphics[width=3in]{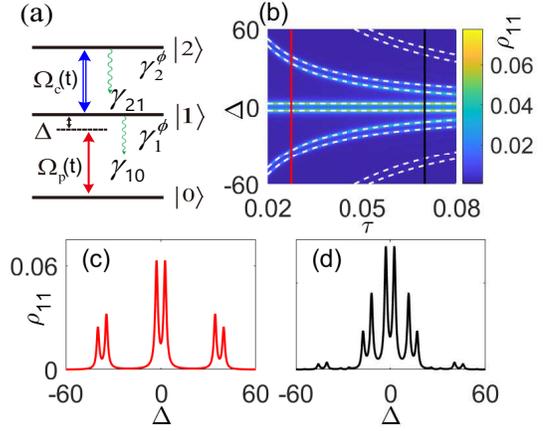}
\caption{(a) Schematic of a three-level system. Based on the above two-level system, a control-field resonantly couples levels $|1\rangle$ and $|2\rangle$ with a periodically modulated Rabi frequency $\Omega_c(t)=\Omega_c(t+\tau)$. $\gamma_{21}$ is the population damping rate from level $|2\rangle$ to $|1\rangle$. $\protect\gamma_{2}^{\protect\phi}$ is the dephasing rate of state $|2\rangle$. (b) A contour map of $\protect\rho_{11}$ as a function of $\Delta$ and $\protect\tau$. The graph is obtained by numerically calculating the Lindblad master equation Eq.~(\ref{eq:threelme}) for $\protect\gamma _{10}=1$, $\protect\gamma _{21}=1.4$, $\protect\gamma_{1}^{\protect\phi}=0.4$, $\protect\gamma_{1}^{\protect\phi}=0.2$ $\Omega_c=10.8$ and $\Omega_p=1$ [i.e., the vertical yellow dashed line in Fig.~\ref{fig:modtwo} (b)]. The white dashed lines show the positions of the resonance peaks, i.e., $(\Delta\pm\Omega_{c}/4)\protect\tau=2 n \protect\pi,~~(n=0, 1, 2\cdots)$. (c), (d) The steady excitation probability $\protect\rho_{11}$ as a function of the probe field detuning with $\tau=0.027$ [i.e., the vertical red line in (b)] and $\tau=0.695$ [i.e., the vertical black line in (b)]. Here we normalized the parameters in terms of $\gamma_{10}$}.
\label{fig:modthree}
\end{figure}
    In this section, we further explore the QIEs in a Floquet three-level system, as shown in Fig.~\ref{fig:modthree}(a). Here we assume the modulation scheme of $\Omega_c(t)$ is asynchronous to that of $\Omega_p(t)$, i.e.,
\begin{eqnarray}  \label{eq:omec}
\Omega_{c}(t)=\left\{
\begin{array}{ll}
0, & ~~~~~t\in[n\tau,(n+\frac{1}{2})\tau] \\
\Omega_c, & ~~~~~t\in[(n+\frac{1}{2})\tau,(n+1)\tau]%
\end{array}
\right.
\end{eqnarray}
with $n=0,1,2\cdots$. Similar to the above two-level system, the Lindblad master equation can be written as
\begin{small}
\begin{eqnarray}\label{eq:threelme}
\dot{\rho}&=&-i[H(t),\rho]+\sum_{j=1,2}\frac{\gamma _{j,j-1}}{2}(2\sigma
_{j-1,j}\rho \sigma _{j,j-1}-\sigma _{jj}\rho -\rho \sigma _{jj})\nonumber\\
&+&\sum_{j=1,2}\gamma _{j}^{\phi }(2\sigma _{jj}\rho \sigma _{jj}-\sigma
_{jj}\rho -\rho \sigma _{jj}),
\end{eqnarray}
\end{small}
with
\begin{eqnarray}\label{eq:threeh}
H(t)&=&\frac{\Delta}{2} (-|0\rangle \langle 0|+|1\rangle\langle 1|+|2\rangle \langle 2|)\nonumber\\
&-&\left[\frac{\Omega _{p}(t)}{2}|0\rangle \langle1|+\frac{\Omega _{c}(t)}{2}|1\rangle \langle 2|+h.c.\right].
\end{eqnarray}
Similar to Eq.~(\ref{eq:two}), here we apply RWA by assuming $\omega=1/\tau\ll 2\omega_{10},2\omega_{21}$, $\Omega_p\ll\omega_{10}$ and $\Omega_c\ll\omega_{21}$.

To study the steady-state properties of the Floquet three-level system, only the data at the end of each modulation period (i.e., the time evolution step is $\tau$) is observed. The dynamics start from the ground state $|0\rangle$, we show the contour map of $\protect\rho_{11}$ as a function of the probe field detuning $\Delta$ and the modulation period $\protect\tau$ in Fig.~\ref{fig:modthree}(b). One finds that each peak in Fig.\ref{fig:modtwo}(b) splits into two when an added control field couples $|1\rangle $ to $|2\rangle $, resulting in multiple transparency windows, which is similar to the multi-chromatic ATS with multi-tone control fields studies in the un-modulated systems~\cite{2006Bichromatic, PhysRevA.57.1355, PhysRevA.68.063810}. The positions of the peaks, as the white dashed lines shown in Fig.~\ref{fig:modthree}(b), could be demonstrated to be exactly the maximal constructive interference of transitions $|+\rangle\leftrightarrow|0\rangle$ and $|-\rangle\leftrightarrow|0\rangle$, which could be achieved by calculating relative phases $\phi _{0\pm }$ in one modulation period. Here $|\pm \rangle =(|2\rangle \pm |1\rangle )/\sqrt{2}$ is the dressed levels induced by the strong control field. During the controlled half-period, the system is in the dressed basis $|\pm \rangle $ with eigenenergy $(\Delta \pm \Omega _{c})/2$. The relative phases accumulated in
the first half period are $\phi _{0\pm }^{\prime }=(\Delta \pm \Omega
_{c}/2)\tau /2$. During the non-controlled half period,
the system is in bare basis. Driven by a probe field with detuning $\Delta $%
, the relative phases $\phi _{0\pm }^{\prime \prime }$ accumulated are $\phi
_{0+}^{\prime \prime }=\phi _{0-}^{\prime \prime }\approx \Delta \tau /2,$
where $\Omega _{p}\ll \Delta $ is assumed. Then the total relative
phases in one period are $\phi _{0\pm }=\phi _{0\pm }^{\prime }+\phi
_{0\pm }^{\prime \prime }=(\Delta \pm \Omega _{c}/4)\tau $. The maximal
interference requires $\phi _{0\pm }=2n\pi $, $~(n=0,1,2\cdots )$.
Therefore, the constraints for absorption peaks are several curves $\Delta
=2n\pi \omega \pm \Omega _{c}/4$, as the white dashed lines shown in Fig.~\ref{fig:modthree}(b). The multiple transparency windows may have promising applications in quantum memory~\cite{PhysRevLett.120.183602}, ground-state cooling~\cite{YiGround} and so on.

Figure.~\ref{fig:modthree}(c) and (d) show the multiple transparency windows more clearly. One finds the asymmetry of the transparency windows except for the central one and the multiple transparency windows move closer to the center (i.e., $\Delta=0$) as $\tau $ increases. Moreover, the central transparency window (CTW) is similar to the traditional EIT and ATS, but the modulation period as a new adjustable dimension enriches the interference properties of the CTW.

For the traditional EIT and ATS in the un-modulated systems, it is widely acknowledged that: ATS exhibits a wide non-interference transparency window between large Stark splittings induced by a strong field \cite{Peng2014EIT, Ling2015Discerning}, while EIT exhibits a narrow transparency window induced by destructive quantum interference (QI) between the Stark splittings~\cite{RevModPhys.77.633}. In the weak probe field approximation, according to Ref.~\cite{2016Interference} the absorption of the probe field is determined by the first-order coherence of the probe transition and is given as
\begin{equation}
\text{Im}(\rho _{10})_{\text{QI}}=\frac{\Omega _{p}}{4B}[(\Delta -\frac{%
\Omega _{c}}{2})^{2}\Gamma +(\Delta +\frac{\Omega _{c}}{2})^{2}\Gamma +2A]
\label{eq:ati}
\end{equation}%
where $A=-[(\frac{\Omega _{c}}{2})^{2}-\Delta ^{2}]\Lambda +(\Gamma -\Lambda)^{2}(\Gamma +\Lambda )$ and $B=[(\Delta +\frac{\Omega _{c}}{2})^{2}+\Gamma^{2}][(\Delta -\frac{\Omega _{c}}{2})^{2}+\Gamma ^{2}]-2[(\frac{\Omega _{c}}{%
2})^{2}-\Delta ^{2}+\Gamma ^{2}]\Lambda ^{2}+\Lambda ^{4}$ with $\Gamma =(\gamma _{10}+\gamma_{21})/4+(\gamma _{1}^{\phi }+\gamma _{2}^{\phi })/2$ and $\Lambda =(\gamma_{10}-\gamma _{21})/4+(\gamma _{1}^{\phi }-\gamma _{2}^{\phi })/2$. Here $\Lambda$ induces quantum interference between the transitions $|+\rangle\leftrightarrow|0\rangle$ and $|-\rangle\leftrightarrow|0\rangle$ (more detail see Appendix \ref{app:dress}). Then according to the well-known ATS (i.e., non-interference) \cite{Liu:20}, $\Lambda$ should be zero and Eq.~(\ref{eq:ati}) reduces to
\begin{equation}
\text{Im}(\rho _{10})_{\text{ATS}}=\frac{\Gamma \Omega _{p}/4}{(\Delta -%
\frac{\Omega _{c}}{2})^{2}+\Gamma ^{2}}+\frac{\Gamma \Omega _{p}/4}{(\Delta +%
\frac{\Omega _{c}}{2})^{2}+\Gamma ^{2}}.  \label{eq:ats}
\end{equation}%
Equation~(\ref{eq:ats}) is exactly the sum of two Lorentzian peaks corresponding to transitions $|+\rangle\leftrightarrow|0\rangle$ and $|-\rangle\leftrightarrow|0\rangle$ with Stark splitting $\Omega _{c}$.
By comparing the values of $\mathrm{Im}(\rho _{10})_{\mathrm{ATS}}$ and $\mathrm{Im}(\rho _{10})_{\mathrm{QI}}$ in
Eqs.~(\ref{eq:ati}) and (\ref{eq:ats}) under $\Delta=0$, we find that when $0<\Lambda <\Gamma
$ ($-\Gamma <\Lambda <0$), it induces destructive (constructive) interference and shallows (deepens) the absorption
valley. Moreover, when $\Lambda \approx \Gamma $, absorption is almost
completely suppressed (i.e., $\mathrm{Im}(\rho _{10})_{\mathrm{QI}}|_{\Delta =0}\approx 0$) due to complete destructive
interference between transitions $|+\rangle\leftrightarrow|0\rangle$ and $|-\rangle\leftrightarrow|0\rangle$. Similarly, one has almost complete constructive interference
for $\Lambda \approx -\Gamma $. When the control field is getting weaker
or the decoherence rates are getting greater, the absorption dip
corresponding to the transparency might disappear. Therefore, the additional condition
for the probe transparency dip to be observed is $\partial \mathrm{Im}(\rho
_{10})_{\mathrm{QI}}/\partial \Delta ^{2}|_{\Delta =0}>0$, giving $\Omega
_{c}>2\sqrt{(\Gamma -\Lambda )^{3}/(3\Gamma -\Lambda )}$.

To explore the interference properties of the CTW in our Floquet three-level system, the AIC method~\cite{PhysRevLett.107.163604} is used to discern the CTW from QI and ATS models in Eqs.~(\ref{eq:ati}) and (\ref{eq:ats}) by evaluating their relative AIC weights for different modulation periods.

\subsection{ATS-like profile}\label{sec:ats}
\begin{figure}[tph]
\includegraphics[width=3in]{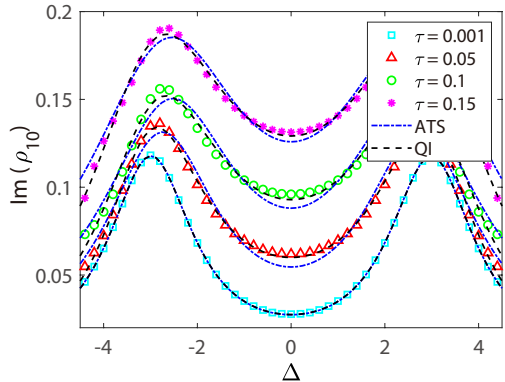}
\caption{Absorption $\mathrm{Im}(\protect\rho _{10})$ as a function of detuning $\Delta $, and various $\protect\tau $ with best fits to
ATS($\Omega _{c},\Omega _{p}$,$\Gamma $) (black dashed lines) and QI($\Omega
_{c},\Omega _{p}$,$\Gamma $,$\Lambda $) (blue dot dashed lines) models
calculated for $\protect\tau =0.001$ (Bright cyan squares) with a good fit
to ATS$(5.391,0.4543,0.9958)$ as well as QI$(5.396,0.4538,0.9949,-0.0071)$, $%
\protect\tau =0.05$ (red triangles) with a better fit to QI$%
(5.69,0.56,1.17,-0.73)$ than ATS$(4.972,0.6535,1.326)$, $\protect\tau =0.1$
(green circles) with a better fit to QI$(5.626,0.7424,1.447,-0.8687)$ than
ATS$(4.641,0.8731,1.607)$, $\protect\tau =0.15$ (Magenta asterisks) with a
better fit to QI$(5.847,1.011,1.657,-0.9817)$ than ATS$(4.763,1.231,1.89)$. Here we normalized the parameters in terms of $\gamma_{10}$}.
\label{fig:atwo}
\end{figure}

In this section, we use the parameters that satisfy the ATS model in Eqs.~(\ref{eq:ats}), i.e., $\gamma _{10}=1$, $\gamma_{21} =1.4$, $\gamma _{1}^{\phi } =0.4$, $\gamma _{2}^{\phi} =0.2$, $\Omega_p=1$ and $\Omega _{c}=10.8$, which is also consistent with the parameters in Fig.~\ref{fig:modthree}. Then $\Gamma =0.9$ and $\Lambda =0$, where a non-interference transparency window appears for the un-modulated system. In Fig.~\ref{fig:atwo}, we plot the numerical results of $\mathrm{Im}(\rho_{10})$ obtained from Eq.~(\ref{eq:threelme}) for the Floquet three-level system. Note that here we only focus on the properties of the CTW and the numerical results is obtained by calculating the dynamic steady states of the Floquet three-level system with time evolution step $\tau$. As an objective way to identify the more appropriate model for the CTW in the Floquet three-level system, we also plot the fitting profile for each $\tau$ using QI($\Omega_{c},\Omega _{p}$,$\Gamma $,$\Lambda $) and  ATS($\Omega _{c},\Omega _{p}$,$\Gamma $) models with fitting parameters $\Omega _{p}$, $\Omega _{c}$, $\Gamma $ and $\Lambda $ in Eqs.~(\ref{eq:ati}) and (\ref{eq:ats}) as a comparison. For small Floquet period, e.g. $\tau=0.001$, ATS and QI fitting
curves merge and both fit well with the simulated absorption. The fitted
value of the interference term $\Lambda$ is close to 0, which indicates that
the CTW in rapidly modulated system has the similar properties to the
transparency in it un-modulated counterparts. This could be understood as
the quantum Zeno effect~\cite{PhysRevA.41.2295}. For larger Floquet periods,
such as $\tau=0.05,0.1,0.15$, we find that the CTW fits better with the QI
model. The fitted interference $\Lambda<0$, which indicates constructive
interference in the CTW. This could be understood qualitatively as that the
periodic driving heat the system and effectively change the decay rates.
However, as the modulation frequency $\omega$ increases, the heating
speed becomes slower, which is known as Floquet prethermalization~\cite%
{PhysRevX.10.021044}. To conclude, in the ATS's parameter regime, the CTW is
ATS-like but with significant interference adjustable in a wide range of modulation
periods, which differentiates the CTW in the Floquet three-level system from the traditional ATS.

The above observations of the CTW's underlying mechanism are supported by
quantitative criteria. To be specific, we apply the AIC method, which uses the relative entropy to identify the most
information model\cite{Peng2014EIT}. The information loss of a given model
with $k$ fitting parameters to the numerical data is quantified by $I=N%
\mathrm{ln}(R/N)+2k$, where $N$ is the number of numerical data for fitting
and $R$ denotes the fitting residual sum of squares. The per-point AIC
contribution is $\bar{I}=I/N$. Hence the AIC per-point weights are
\begin{equation}
\bar{w}_{QI}=\frac{\mathrm{exp}(-\frac{1}{2}\bar{I}_{QI})}{\mathrm{exp}(-%
\frac{1}{2}\bar{I}_{ATS})+\mathrm{exp}(-\frac{1}{2}\bar{I}_{QI})}
\label{eq:ataic}
\end{equation}
for QI model and $\bar{w}_{\mathrm{ATS}}=1-\bar{w}_{QI}$ for ATS model.
Greater weight means more likelihood fitting. For $\tau
=0.001,0.05,0.1,0.15 $, $\bar{w}_{QI}=0.51,0.74,0.75,0.64$ respectively. Not
surprisingly, for $\tau =0.001$, $\bar{w}_{QI}\approx \bar{w}_{ATS}$. In
general regions of $\tau $, $\bar{w}_{QI}>\bar{w}_{ATS}$, which means that
the QI model is indeed the more appropriate model for the CTW in the Floquet three-level system.

\subsection{EIT-like profile}\label{sec:eit}
\begin{figure}[pth]
\includegraphics[width=3.2in]{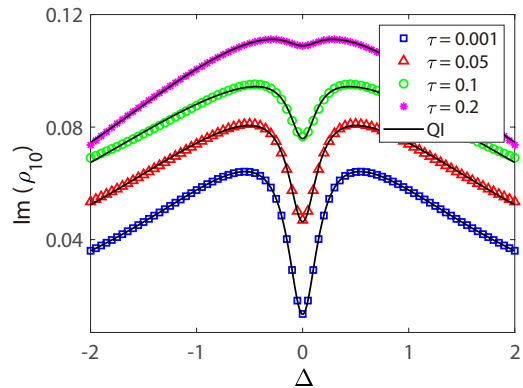}\caption{Absorption $\mathrm{Im}%
~(\rho_{10})$ as a function of detuning $\Delta$, and various $\tau$
with best fits to QI($\Omega_{c},\Omega_{p}$,$\Gamma$,$\Lambda$) (black solid lines) model calculated for $\tau=0.001, 0.05,0.1,0.2$ with a good fit to
QI$(0.4809,1.812,1.875,1.817)$, QI$(0.7392,1.474,2.31,2.155)$, QI$(1.018, 1.092,2.736,2.529)$ and QI(1.109,0.5408,2.62,2.265) respectively. Here we normalized the parameters in terms of $\gamma_{10}$}.%
\label{fig:meit}%
\end{figure}
In this section, we further consider the parameters that satisfy the QI model in Eqs.~(\ref{eq:ati}), i.e., $\gamma_{10}=1$, $\gamma_{21}=0.1$, $\gamma_{1}^{\phi}=3$, $\gamma_{2}^{\phi}=0$, $\Omega_p=1$ and $\Omega_{c}= 3.55$. Then $\Gamma=1.775,\Lambda=1.725$, which corresponds to almost completely destructive interference between transitions $|+\rangle\leftrightarrow|0\rangle$ and $|-\rangle\leftrightarrow|0\rangle$ under $\Delta=0$. Similar to Sec.~\ref{sec:ats}, Fig.~\ref{fig:meit} shows the numerical results of $\mathrm{Im}(\rho_{10})$ and the fitting profile for each $\tau$ using the QI model in Eqs.~(\ref{eq:ati}). Note that here the poor ATS fitting results are ignored. When $\tau=0.001$, the CTW has a narrow dip at zero detuning, which is similar to the conventional EIT in the un-modulated system. As the modulation period increases, the dip becomes shallow, and the modulation weakens the strength of the destructive interference. Therefore, in the EIT parameter regime, the CTW is demonstrated to exhibit an EIT-like profile but with interference adjustable with the modulation period.

\section{Discussions and conclusions}
We investigate the QIEs in the Floquet two- and three-level systems. In the Floquet two-level system, the monochromatic periodic pulses (e.g. square-wave) generate equivalent poly-chromatic drivings, which enable lotus-like multi-peak phenomenon and the coherent destruction of tunneling effect. In the Floquet three-level system, where the probe and control fields are asynchronously modulated by the same square-wave pulses, multiple transparency windows are observed and the quantum interference of the CTW can be tuned by the modulation periods of the external fields. And the CTW becomes EIT-like or ATS-like by adjusting the modulation periods of the external fields without changing the properties of the systems, which will greatly improve the application prospects of the existing systems. Moreover, the multiple transparency windows may provide a powerful platform beyond the applications based on the traditional single transparent
window, such as multi-frequency all-optical switching, which can switch on/off multi-chromatic fields simultaneously.

The modulation scheme proposed here can be easily implemented experimentally
in various three-level systems, such as atoms gases~\cite%
{PhysRevLett.79.2959,NovikovaElectromagnetically}, superconducting quantum
circuits~\cite{You2011Atomic}, quantum dots~\cite%
{XuCoherent}, nanoplasmonics~\cite{LiuPlasmonic},
optomechanics~\cite{safavi2011, JiangElectromagnetically} and so on. For a
qutrit, which is a three-level artificial atom in the superconducting
circuits, it can be manipulated by microwave fields. The modulation period $\tau$ that can be realized experimentally is about dozens of ns~\cite{HanPRApplied}, which is much smaller than the coherence time of the system (about 0.5ms)~\cite{ WOS:000742352300001}. For such systems,
hundreds of periods can be realized within the coherence time of the system.

\begin{acknowledgements}
This work is supported by the National Natural Science
Foundation of China under Grant Nos. U1930201, 91836101, U1801661 and 11934010, the
Key-Area Research and Development Program of Guangdong Province (Grant No.
2018B030326001), Guangdong Provincial Key Laboratory(Grant No.
2019B121203002), the Science, Technology and Innovation Commission of
Shenzhen Municipality (No. JCYJ20170412152620376, No.
KYTDPT20181011104202253), Grant No.2016ZT06D348, and the Natural Science
Foundation of Guangdong Province (Grant No. 2017B030308003). X.-Q.L acknowledges the support by the Foundation of Hunan Educational Committee (Grant No. 19C1585), the Natural Science Foundation of Hunan Province of China (Grant No. 2020JJ5466), and the National Natural Science Foundation of China (Grants No. 12104214).
\end{acknowledgements}

\appendix

\section{The validity of rotating-wave approximation for time-dependent Hamiltonian}\label{app:rwa}
\begin{figure}[pth]
\centering
\includegraphics[width=3.2in]{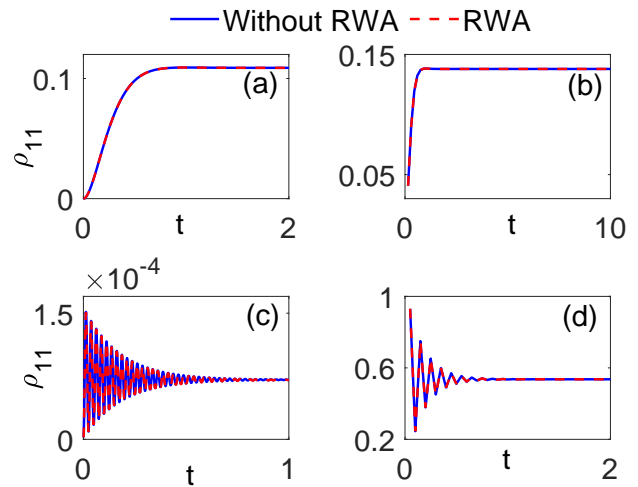}
\caption{ Comparison of numerical results of $\rho_{11}$ without RWA [obtained from Eq.~(\ref{eq:twonorwa})] and with RWA [obtained from Eq.~(\ref{eq:tworwa})]  for (a) $\tau=0.001$, $\Delta=0$, $\Omega_p=1$, (b) $\tau=0.15$, $\Delta=0$, $\Omega_p=1$, (c)$\tau=0.001$, $\Delta=40$, $\Omega_p=1$, (d)$\tau=0.05$, $\Delta=0$, $\Omega_p=200$. Other parameters are
$\omega_p=6000$, $\omega_{10}=\Delta+\omega_p$, $\gamma_1^{\phi}=0.4$, $\gamma_{10}=1$. Here we normalized the parameters in terms of $\gamma_{10}$}.

\label{fig:valid1}
\end{figure}

In this section, we verify the validity of RWA with the Floquet two-level system described in the main text. The exact Hamiltonian is
\begin{eqnarray}\label{eq:twonorwa}
   H(t)= H_0-[\frac{\Omega _{p}(t)(1+e^{2i\omega_pt})}{2}|0\rangle \langle1|+h.c.]
\end{eqnarray}
with $H_0=\Delta (-|0\rangle \langle 0|+|1\rangle \langle1|)/2$.
Here exp$(2i\omega_pt)$ is the counter-rotating term proportional to exp$[\pm i(\omega_{10}+\omega_p-\Delta)t]$.
Substituting Eq.~(\ref{eq:foueop}) in the main text into Eq.~(\ref{eq:twonorwa}), the coupling term
between levels $|0\rangle$ and $|1\rangle$ becomes
\begin{eqnarray}\label{eq:rwava}
  \left[\frac{\Omega _{p}}{2}-\sum_{n=1}^{\infty }(-1)^{n}\Omega_{pn}\cos (\omega _{n}t)\right](1+e^{2i\omega_pt}).
\end{eqnarray}
The counter-rotating term exp$(2i\omega_pt)$ causes rotating terms proportional to exp$[\pm i(\omega_{n}\pm 2\omega_p)t]$, which is usually a fast rotating term ($2\omega_p\gg\omega_n$) that can be ignored. For example in superconducting circuit (atom) experiments, the frequency of near-resonant microwave (laser) is several GHz (THz)~\cite{PhysRevLett.107.240501,PhysRevB.86.100506}, however, the modulation frequency is about dozens of MHZ~\cite{HanPRApplied}. Another condition for RWA to be applicable in Eq.~(\ref{eq:rwava}) is $\Omega_{pn}\ll\omega_n+\omega_p$, which can be simplified to $\Omega_p\ll\omega_p\approx\omega_{10}$, and this is the scenario we considered. Note that here we focus on the situation for small n, because the corresponding Rabi frequency $\Omega_{pn}$ decreases sharply as n increases. Therefore, in our Floquet system, the RWA is validly and in $\omega_p$'s rotating frame, Eq.~(\ref{eq:twonorwa}) becomes
\begin{eqnarray}\label{eq:tworwa}
   H(t)_{\text{RWA}}= H_0-[\frac{\Omega _{p}(t)}{2}|0\rangle \langle1|+h.c.].
\end{eqnarray}
In Fig.~\ref{fig:valid1}, we compare the numerical results of $\rho_{11}$ obtained from Eq.~(\ref{eq:twonorwa}) and Eq.~(\ref{eq:tworwa}) for the extensive parameters that we study in the main text, and find that the RWA results agree well with the exact results. Similarly, for our Floquet three-level system, the conditions for the RWA to be applicable are $\omega\ll 2\omega_{10},2\omega_{21}$, $\Omega_p\ll\omega_{10}$ and $\Omega_c\ll\omega_{21}$.

\section{Analytical results of the Floquet two-level system}\label{app:twor}

\begin{figure}[pth]
\includegraphics[width=3.2in]{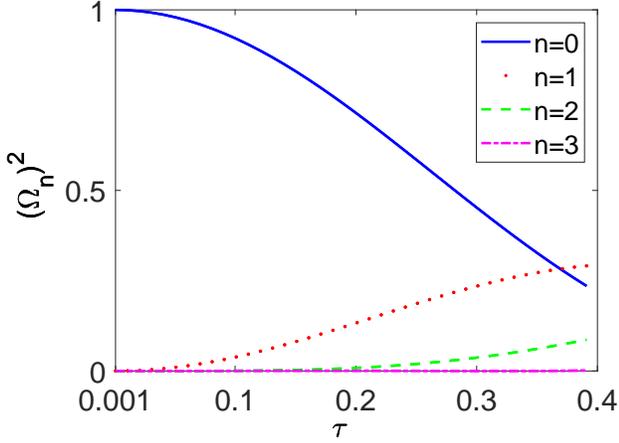}\caption{$(\Omega_n)^2$ [obtained from Eq.~(\ref{eq:omega})] as a function of $\tau$ with $\Omega_p=1$. Here we normalized the parameters in terms of $\gamma_{10}$}.%
\label{fig:omega}
\end{figure}
In the weak driving range ($\Omega_{p}<\omega$), the sidebands are well separated and show clearly resonance peaks caused by the different frequency components. The
analytical results of these sidebands can be easily obtained as
\begin{eqnarray}\label{eq:tows}
\rho _{11} &=&\frac{\frac{\gamma'_1 }{2\gamma _{10}}(\frac{\Omega _{p}}{2})^{2}}{(\gamma'_1)^2+\Delta ^{2}+\frac{\gamma'_1}{\gamma _{10}}(\frac{\Omega _{p}}{2})^{2}}\nonumber\\
&&+\sum_{n=1}^{\infty }\frac{\frac{\gamma'_1}{2\gamma _{10}}\Omega _{pn}^{2}}
{(\gamma'_1)^{2}+(\Delta +\omega _{n})^{2}+\frac{\gamma'_1}{\gamma _{10}}\Omega_{pn}^{2}}
\end{eqnarray}
with $\gamma _{1}^{\prime }=\gamma _{10}/2+\gamma_{1}^{\phi }$. Equation~(\ref{eq:tows}) consists of a series of independent Lorentizians,
each one with width $\sqrt{(\gamma'_1)^{2}+\gamma'_1\Omega _{pn}^{2}/\gamma _{10}}$.

In the strong driving range ($\Omega_{p}>\omega$), the pattern becomes too
complex to get the analytic solution. However, we can get the analytical
expression for the resonance condition ($\triangle=0$) and expand to the
general case. When $\triangle=0$, the optical Bloch equations are
\begin{eqnarray}
\frac{\partial V (t)}{\partial t}&=& -\gamma'_1 V, \\
\frac{\partial U (t)}{\partial t} &=& \frac{\gamma_{10}}{2}-\frac{1}{2}
(\gamma_{10}+\gamma'_1) U(t)+i\Omega_p(t)U(t)\nonumber\\
&-&\frac{1}{2}(\gamma_{10}-\gamma'_1)W(t) ,  \label{eq:obu} \\
\frac{\partial W (t)}{\partial t} &=& \frac{\gamma_{10}}{2}-\frac{1}{2}
(\gamma_{10}+\gamma'_1)W(t)-i\Omega_p(t)W(t) \nonumber\\
&-&\frac{1}{2}(\gamma_{10}-\gamma'_1)U(t), \label{eq:obw}
\end{eqnarray}
where
\begin{eqnarray}  \label{eq:topi}
V(t)&=&\frac{1}{2}\left[\rho_{10}(t)+\rho_{01}(t)\right], \\
U(t)&=&\frac{1}{2}[\rho_{11}(t)-\rho_{00}(t)+\rho_{01}(t)-\rho_{10}(t)],
\label{eq:urho} \\
W(t)&=&\frac{1}{2}[\rho_{11}(t)-\rho_{00}(t)-\rho_{01}(t)+\rho_{10}(t)].
\end{eqnarray}
In the absence of damping ($\gamma_{10}=\gamma'_1=0$), the solutions of Eqs.~(\ref{eq:obu})and (\ref{eq:obw}) show that the components $U(t)$ and $%
W(t)$ oscillate with frequencies $\pm \Omega_p(t)$, and their
oscillation frequencies differ by $2\Omega_p(t)$. Therefore, in a frame
oscillating with $\Omega_p(t)$ the terms proportional to $%
(\gamma_{10}-\gamma'_1)/2$ oscillate with $\pm 2\Omega_p(t)$. Thus, we can
ignore the rapidly oscillating terms, and obtain the solutions of $U(t)$ and
$W(t)$ by direct integration~\cite{Ficek2000Saturation},
\begin{eqnarray}  \label{eq:solu}
U(t) = -\frac{\gamma_{10}}{2}\int_{0}^{t}dt^{\prime A(t-t^{\prime
})}e^{\sum\limits_{n=1}^{\infty}B[\sin(\omega_nt^{\prime
})-\sin(\omega_nt)]},  \nonumber \\
\end{eqnarray}
where
\begin{eqnarray}
  A = -\frac{1}{2}(\gamma_{10}+\gamma'_1-i\Omega_p),~~
  B = \frac{i(-1)^{n+1}\Omega_{pn}}{\omega_n}.
\end{eqnarray}
It is seen from Eq.~(\ref{eq:urho}) that
\begin{eqnarray}
\frac{\rho_{11}(t)-\rho_{00}(t)}{2}=\mathrm{Re}[U(t)],
\end{eqnarray}
Combined with $\rho_{11}(t)+\rho_{00}(t)=1$, the population of state $%
|1\rangle$ is
\begin{equation}  \label{eq:rho11t}
\rho_{11}(t)=\frac{1}{2}-\mathrm{Re}[U(t)].
\end{equation}
From Eq.~(\ref{eq:urho}), we also get

\begin{equation}\label{eq:rho10}
\mathrm{Im}[\rho_{10}(t)]=-\mathrm{Im}[U(t)].
\end{equation}

Substituting Eq.~(\ref{eq:solu}) into Eq.~(\ref{eq:rho11t}) and Eq.~(\ref{eq:rho10}), and
straightforward calculating the steady solution of $\rho_{11}$ and $\mathrm{Im}(\rho_{10})$
\begin{eqnarray}  \label{eq:epss1}
\rho_{11}(\infty)&=&\frac{1}{2}-\frac{\gamma_{10}\gamma_1}{2}%
\sum_{n=-\infty}^{\infty}\frac{\Omega_n^2}{\gamma_1^ 2
+(\Omega_{p}/2-n\omega)^2},  \nonumber \\ \\
\mathrm{Im}[\rho_{10}(\infty)]&=&\frac{\gamma_{10}}{2}%
\sum_{n=-\infty}^{\infty}\frac{\Omega_n^2(\Omega_{p}/2-n\omega)}{\gamma_1^ 2
+(\Omega_{p}/2-n\omega)^2},
\end{eqnarray}
where
\begin{eqnarray}\label{eq:omega}
\Omega_n
&=&\sum_{\xi=-\infty}^{\infty}\sum_{l=-\infty}^{\infty}\cdots\sum_{g=-%
\infty}^{\infty} J_{\xi}\bigg(\frac{-2\Omega_{p}}{\omega\pi}\bigg)J_{l}\bigg(%
\frac{2\Omega_{p}}{9\omega\pi}\bigg)\cdots \nonumber\\
&&J_{g}\bigg(\frac{(-1)^{q}2\Omega_{p}}{(2q-1)^{2}\omega\pi}\bigg),
\end{eqnarray}
with $\xi=n-3l-\cdots-(2q-1)g $ and $\gamma_1=(\gamma'_1+\gamma_{10})/2$. From Eqs.~(\ref{eq:epss1})-(\ref{eq:omega}), we find that  the steady solutions of $\rho_{11}$ and $\mathrm{Im}(\rho_{10})$ are related with $\Omega_n^2$, which is a series of Bessel functions with variable $\Omega_p/\omega$. In fig.~\ref{fig:omega}, we plot $(\Omega_n)^2$ as a function of $\tau$ with $\Omega_p=1$ and find that only $\Omega_0^2\neq 0$, when $\tau\rightarrow 0$.

\section{Derivation of QI and ATS models in Eqs.~(\ref{eq:ati}) and (\ref{eq:ats})}\label{app:dress}
In this section, we derive the QI and ATS models in Eqs.~(\ref{eq:ati}) and (\ref{eq:ats}) with an un-modulated Hamiltonian
\begin{equation}\label{eq:threeun}
H=\frac{\Delta}{2} (|1\rangle\langle 1|+|2\rangle \langle 2|-|0\rangle \langle 0|)-(\frac{\Omega _{p}}{2}|0\rangle \langle1|+\frac{\Omega _{c}}{2}|1\rangle \langle 2|+h.c.)
\end{equation}
The dynamics affected by the control field could be better studied in dressed state representation by replacing the $H(t)$ in Eq.~(\ref{eq:threelme}) with Eq.~(\ref{eq:threeun})~\cite{2016Interference}. The elements in the density matrix
associated with the probed transition dynamics are
\begin{eqnarray}
\dot{\rho}_{-0} &=&-\left[i(\Delta +\frac{\Omega _{c}}{2})+\Gamma \right]\rho_{-0}-\Lambda \rho _{+0}\nonumber \\
&+&\frac{i\Omega _{p}}{2\sqrt{2}}(\rho _{00}-\rho _{--}-\rho _{-+}),\label{eq:rho-0} \\
\dot{\rho}_{+0}& =&-\left[i(\Delta -\frac{\Omega _{c}}{2})+\Gamma \right]\rho_{+0}-\Lambda \rho _{-0}\nonumber\\
&+&\frac{i\Omega _{p}}{2\sqrt{2}}(\rho _{00}-\rho _{++}-\rho _{+-})
\label{eq:rho+0}.
\end{eqnarray}%
From Eqs.~(\ref{eq:rho-0}) and (\ref{eq:rho+0}), one finds that $\Lambda$ cross-couples the two dressed states' dynamics and induces quantum interference between $\rho_{+0}$ and $\rho_{-0}$. Then we give the first-order steady solutions of $\rho_{+0}$ and $\rho_{-0}$ with the weak probe field approximation, where the steady-state zero-order populations
and coherence are $\rho^0_{00}=1,~\rho^0_{++}=\rho^0_{--}=\rho^0_{+-}=0$. With these conditions, Eqs.~(\ref{eq:rho-0}) and (\ref{eq:rho+0}) become
\begin{eqnarray}
\dot{\rho}_{-0} &  =-[i(\Delta+\frac{\Omega_{c}}{2})+\Gamma]\rho_{-0}%
-\Lambda\rho_{+0}
  +\frac{i\Omega_{p}}{2\sqrt{2}},\label{eq:rho-01}\\
\dot{\rho}_{+0} &  =-[i(\Delta-\frac{\Omega_{c}}{2})+\Gamma]\rho_{+0}%
-\Lambda\rho_{-0}
 +\frac{i\Omega_{p}}{2\sqrt{2}}.\label{eq:rho+01}%
\end{eqnarray}
This set of equations can be solved by writing in the matrix form,
\begin{equation}
\dot{R}=-MR+A
\end{equation}
with
\begin{eqnarray}
R&=&\left(
\begin{array}{c}
\rho_{-0}  \\
 \rho_{+0}
\end{array}
\right),\\
M&=&\left(
\begin{array}{cc}
i(\Delta+\frac{\Omega_{c}}{2})+\Gamma &\Lambda  \\
 \Lambda    & i(\Delta-\frac{\Omega_{c}}{2})+\Gamma
\end{array}
\right),\\
A&=&\left(
\begin{array}{c}
 \frac{i\Omega_{p}}{2\sqrt{2}} \\
 \frac{i\Omega_{p}}{2\sqrt{2}}
\end{array}
\right),
\end{eqnarray}
and then integrating
\begin{eqnarray}
R(t)=\int_{-\infty}^t e^{-M(t-t')Adt'}=M^{-1}A.
\end{eqnarray}
This yields
\begin{eqnarray}
\rho_{-0}&=&\frac{i\Omega_p}{2\sqrt{2}}\frac{i(\Delta-\frac{\Omega_c}{2})+\Gamma-\Lambda}
{[i(-\Delta-\frac{\Omega_c}{2})-\Gamma][i(-\Delta+\frac{\Omega_c}{2})-\Gamma]-\Lambda^2},\nonumber\\ \\
\rho_{+0}&=&\frac{i\Omega_p}{2\sqrt{2}}\frac{i(\Delta+\frac{\Omega_c}{2})+\Gamma-\Lambda}
{[i(-\Delta-\frac{\Omega_c}{2})-\Gamma][i(-\Delta+\frac{\Omega_c}{2})-\Gamma]-\Lambda^2}.\nonumber\\
\end{eqnarray}
Then the absorption of the probe field is Im$(\rho_{10})=\mathrm{Im}(\rho_{+0}+\rho_{-0})/\sqrt{2}$.


\begin{thebibliography}{62}
\expandafter\ifx\csname natexlab\endcsname\relax\def\natexlab#1{#1}\fi
\expandafter\ifx\csname bibnamefont\endcsname\relax
  \def\bibnamefont#1{#1}\fi
\expandafter\ifx\csname bibfnamefont\endcsname\relax
  \def\bibfnamefont#1{#1}\fi
\expandafter\ifx\csname citenamefont\endcsname\relax
  \def\citenamefont#1{#1}\fi
\expandafter\ifx\csname url\endcsname\relax
  \def\url#1{\texttt{#1}}\fi
\expandafter\ifx\csname urlprefix\endcsname\relax\def\urlprefix{URL }\fi
\providecommand{\bibinfo}[2]{#2}
\providecommand{\eprint}[2][]{\url{#2}}

\bibitem[{\citenamefont{Jing et~al.}(2015)\citenamefont{Jing, Wu, Byrd, You,
  Yu, and Wang}}]{PhysRevLett.114.190502}
\bibinfo{author}{\bibfnamefont{J.}~\bibnamefont{Jing}},
  \bibinfo{author}{\bibfnamefont{L.-A.} \bibnamefont{Wu}},
  \bibinfo{author}{\bibfnamefont{M.}~\bibnamefont{Byrd}},
  \bibinfo{author}{\bibfnamefont{J.~Q.} \bibnamefont{You}},
  \bibinfo{author}{\bibfnamefont{T.}~\bibnamefont{Yu}}, \bibnamefont{and}
  \bibinfo{author}{\bibfnamefont{Z.-M.} \bibnamefont{Wang}},
  \bibinfo{journal}{Phys. Rev. Lett.} \textbf{\bibinfo{volume}{114}},
  \bibinfo{pages}{190502} (\bibinfo{year}{2015}).

\bibitem[{\citenamefont{Zhang et~al.}(2007)\citenamefont{Zhang, Dobrovitski,
  Santos, Viola, and Harmon}}]{PhysRevB.75.201302}
\bibinfo{author}{\bibfnamefont{W.}~\bibnamefont{Zhang}},
  \bibinfo{author}{\bibfnamefont{V.~V.} \bibnamefont{Dobrovitski}},
  \bibinfo{author}{\bibfnamefont{L.~F.} \bibnamefont{Santos}},
  \bibinfo{author}{\bibfnamefont{L.}~\bibnamefont{Viola}}, \bibnamefont{and}
  \bibinfo{author}{\bibfnamefont{B.~N.} \bibnamefont{Harmon}},
  \bibinfo{journal}{Phys. Rev. B} \textbf{\bibinfo{volume}{75}},
  \bibinfo{pages}{201302(R)} (\bibinfo{year}{2007}).

\bibitem[{\citenamefont{Zhang et~al.}(2008)\citenamefont{Zhang, Konstantinidis,
  Dobrovitski, Harmon, Santos, and Viola}}]{PhysRevB.77.125336}
\bibinfo{author}{\bibfnamefont{W.}~\bibnamefont{Zhang}},
  \bibinfo{author}{\bibfnamefont{N.~P.} \bibnamefont{Konstantinidis}},
  \bibinfo{author}{\bibfnamefont{V.~V.} \bibnamefont{Dobrovitski}},
  \bibinfo{author}{\bibfnamefont{B.~N.} \bibnamefont{Harmon}},
  \bibinfo{author}{\bibfnamefont{L.~F.} \bibnamefont{Santos}},
  \bibnamefont{and} \bibinfo{author}{\bibfnamefont{L.}~\bibnamefont{Viola}},
  \bibinfo{journal}{Phys. Rev. B} \textbf{\bibinfo{volume}{77}},
  \bibinfo{pages}{125336} (\bibinfo{year}{2008}).

\bibitem[{\citenamefont{Else et~al.}(2016)\citenamefont{Else, Bauer, and
  Nayak}}]{PhysRevLett.117.090402}
\bibinfo{author}{\bibfnamefont{D.~V.} \bibnamefont{Else}},
  \bibinfo{author}{\bibfnamefont{B.}~\bibnamefont{Bauer}}, \bibnamefont{and}
  \bibinfo{author}{\bibfnamefont{C.}~\bibnamefont{Nayak}},
  \bibinfo{journal}{Phys. Rev. Lett.} \textbf{\bibinfo{volume}{117}},
  \bibinfo{pages}{090402} (\bibinfo{year}{2016}).

\bibitem[{\citenamefont{Sacha and Zakrzewski}(2018)}]{SachaTime}
\bibinfo{author}{\bibfnamefont{K.}~\bibnamefont{Sacha}} \bibnamefont{and}
  \bibinfo{author}{\bibfnamefont{J.}~\bibnamefont{Zakrzewski}},
  \bibinfo{journal}{Rept. Prog. Phys.} \textbf{\bibinfo{volume}{81}},
  \bibinfo{pages}{016401} (\bibinfo{year}{2018}).

\bibitem[{\citenamefont{Oliver et~al.}(2005)\citenamefont{Oliver, Yu, Lee,
  Berggren, Levitov, and Orlando}}]{Oliver2005Mach}
\bibinfo{author}{\bibfnamefont{W.~D.} \bibnamefont{Oliver}},
  \bibinfo{author}{\bibfnamefont{Y.}~\bibnamefont{Yu}},
  \bibinfo{author}{\bibfnamefont{J.~C.} \bibnamefont{Lee}},
  \bibinfo{author}{\bibfnamefont{K.~K.} \bibnamefont{Berggren}},
  \bibinfo{author}{\bibfnamefont{L.~S.} \bibnamefont{Levitov}},
  \bibnamefont{and} \bibinfo{author}{\bibfnamefont{T.~P.}
  \bibnamefont{Orlando}}, \bibinfo{journal}{Science}
  \textbf{\bibinfo{volume}{310}}, \bibinfo{pages}{1653} (\bibinfo{year}{2005}).

\bibitem[{\citenamefont{Shevchenko et~al.}(2010)\citenamefont{Shevchenko,
  Ashhab, and Nori}}]{Shevchenko20101}
\bibinfo{author}{\bibfnamefont{S.}~\bibnamefont{Shevchenko}},
  \bibinfo{author}{\bibfnamefont{S.}~\bibnamefont{Ashhab}}, \bibnamefont{and}
  \bibinfo{author}{\bibfnamefont{F.}~\bibnamefont{Nori}},
  \bibinfo{journal}{Phys. Rep.} \textbf{\bibinfo{volume}{492}},
  \bibinfo{pages}{1} (\bibinfo{year}{2010}).

\bibitem[{\citenamefont{Degert et~al.}(2002)\citenamefont{Degert, Wohlleben,
  Chatel, Motzkus, and Girard}}]{PhysRevLett.89.203003}
\bibinfo{author}{\bibfnamefont{J.}~\bibnamefont{Degert}},
  \bibinfo{author}{\bibfnamefont{W.}~\bibnamefont{Wohlleben}},
  \bibinfo{author}{\bibfnamefont{B.}~\bibnamefont{Chatel}},
  \bibinfo{author}{\bibfnamefont{M.}~\bibnamefont{Motzkus}}, \bibnamefont{and}
  \bibinfo{author}{\bibfnamefont{B.}~\bibnamefont{Girard}},
  \bibinfo{journal}{Phys. Rev. Lett.} \textbf{\bibinfo{volume}{89}},
  \bibinfo{pages}{203003} (\bibinfo{year}{2002}).

\bibitem[{\citenamefont{Han et~al.}(2019)\citenamefont{Han, Luo, Li, Zhang,
  Wang, Tsai, Nori, and You}}]{HanPRApplied}
\bibinfo{author}{\bibfnamefont{Y.}~\bibnamefont{Han}},
  \bibinfo{author}{\bibfnamefont{X.-Q.} \bibnamefont{Luo}},
  \bibinfo{author}{\bibfnamefont{T.-F.} \bibnamefont{Li}},
  \bibinfo{author}{\bibfnamefont{W.}~\bibnamefont{Zhang}},
  \bibinfo{author}{\bibfnamefont{S.-P.} \bibnamefont{Wang}},
  \bibinfo{author}{\bibfnamefont{J.~S.} \bibnamefont{Tsai}},
  \bibinfo{author}{\bibfnamefont{F.}~\bibnamefont{Nori}}, \bibnamefont{and}
  \bibinfo{author}{\bibfnamefont{J.~Q.} \bibnamefont{You}},
  \bibinfo{journal}{Phys. Rev. Applied} \textbf{\bibinfo{volume}{11}},
  \bibinfo{pages}{014053} (\bibinfo{year}{2019}).

\bibitem[{\citenamefont{Wu and An}(2020)}]{PhysRevB.102.041119}
\bibinfo{author}{\bibfnamefont{H.}~\bibnamefont{Wu}} \bibnamefont{and}
  \bibinfo{author}{\bibfnamefont{J.-H.} \bibnamefont{An}},
  \bibinfo{journal}{Phys. Rev. B} \textbf{\bibinfo{volume}{102}},
  \bibinfo{pages}{041119(R)} (\bibinfo{year}{2020}).

\bibitem[{\citenamefont{Bukov et~al.}(2014)\citenamefont{Bukov, D'Alessio, and
  Polkovnikov}}]{2014Universal}
\bibinfo{author}{\bibfnamefont{M.}~\bibnamefont{Bukov}},
  \bibinfo{author}{\bibfnamefont{L.}~\bibnamefont{D'Alessio}},
  \bibnamefont{and}
  \bibinfo{author}{\bibfnamefont{A.}~\bibnamefont{Polkovnikov}},
  \bibinfo{journal}{Adv. Phys.} \textbf{\bibinfo{volume}{64}},
  \bibinfo{pages}{139} (\bibinfo{year}{2014}).

\bibitem[{\citenamefont{Song et~al.}(2020)\citenamefont{Song, You, Xu, Yang,
  and An}}]{PhysRevApplied.14.054049}
\bibinfo{author}{\bibfnamefont{W.-L.} \bibnamefont{Song}},
  \bibinfo{author}{\bibfnamefont{J.-B.} \bibnamefont{You}},
  \bibinfo{author}{\bibfnamefont{J.~K.} \bibnamefont{Xu}},
  \bibinfo{author}{\bibfnamefont{W.~L.} \bibnamefont{Yang}}, \bibnamefont{and}
  \bibinfo{author}{\bibfnamefont{J.-H.} \bibnamefont{An}},
  \bibinfo{journal}{Phys. Rev. Applied} \textbf{\bibinfo{volume}{14}},
  \bibinfo{pages}{054049} (\bibinfo{year}{2020}).

\bibitem[{\citenamefont{Luo et~al.}(2008)\citenamefont{Luo, Xie, and
  Wu}}]{PhysRevA.77.053601}
\bibinfo{author}{\bibfnamefont{X.}~\bibnamefont{Luo}},
  \bibinfo{author}{\bibfnamefont{Q.}~\bibnamefont{Xie}}, \bibnamefont{and}
  \bibinfo{author}{\bibfnamefont{B.}~\bibnamefont{Wu}}, \bibinfo{journal}{Phys.
  Rev. A} \textbf{\bibinfo{volume}{77}}, \bibinfo{pages}{053601}
  (\bibinfo{year}{2008}).

\bibitem[{\citenamefont{Zeng et~al.}(2020)\citenamefont{Zeng, Li, Yang, Xiao,
  and Luo}}]{PhysRevA.102.012221}
\bibinfo{author}{\bibfnamefont{Z.-Y.} \bibnamefont{Zeng}},
  \bibinfo{author}{\bibfnamefont{L.}~\bibnamefont{Li}},
  \bibinfo{author}{\bibfnamefont{B.}~\bibnamefont{Yang}},
  \bibinfo{author}{\bibfnamefont{J.}~\bibnamefont{Xiao}}, \bibnamefont{and}
  \bibinfo{author}{\bibfnamefont{X.}~\bibnamefont{Luo}},
  \bibinfo{journal}{Phys. Rev. A} \textbf{\bibinfo{volume}{102}},
  \bibinfo{pages}{012221} (\bibinfo{year}{2020}).

\bibitem[{\citenamefont{Han et~al.}(2020)\citenamefont{Han, Luo, Li, and
  Zhang}}]{PhysRevA.101.022108}
\bibinfo{author}{\bibfnamefont{Y.}~\bibnamefont{Han}},
  \bibinfo{author}{\bibfnamefont{X.-Q.} \bibnamefont{Luo}},
  \bibinfo{author}{\bibfnamefont{T.-F.} \bibnamefont{Li}}, \bibnamefont{and}
  \bibinfo{author}{\bibfnamefont{W.}~\bibnamefont{Zhang}},
  \bibinfo{journal}{Phys. Rev. A} \textbf{\bibinfo{volume}{101}},
  \bibinfo{pages}{022108} (\bibinfo{year}{2020}).

\bibitem[{\citenamefont{Yudin et~al.}(2016)\citenamefont{Yudin, Taichenachev,
  and Basalaev}}]{PhysRevA.93.013820}
\bibinfo{author}{\bibfnamefont{V.~I.} \bibnamefont{Yudin}},
  \bibinfo{author}{\bibfnamefont{A.~V.} \bibnamefont{Taichenachev}},
  \bibnamefont{and} \bibinfo{author}{\bibfnamefont{M.~Y.}
  \bibnamefont{Basalaev}}, \bibinfo{journal}{Phys. Rev. A}
  \textbf{\bibinfo{volume}{93}}, \bibinfo{pages}{013820}
  (\bibinfo{year}{2016}).

\bibitem[{\citenamefont{Ikeda and Sato}(2020)}]{2020General}
\bibinfo{author}{\bibfnamefont{T.~N.} \bibnamefont{Ikeda}} \bibnamefont{and}
  \bibinfo{author}{\bibfnamefont{M.}~\bibnamefont{Sato}},
  \bibinfo{journal}{Sci. Adv.} \textbf{\bibinfo{volume}{6}},
  \bibinfo{pages}{4019} (\bibinfo{year}{2020}).

\bibitem[{\citenamefont{Han et~al.}(2022)\citenamefont{Han, Zhang, and
  Li}}]{Han:22}
\bibinfo{author}{\bibfnamefont{Y.}~\bibnamefont{Han}},
  \bibinfo{author}{\bibfnamefont{W.}~\bibnamefont{Zhang}}, \bibnamefont{and}
  \bibinfo{author}{\bibfnamefont{W.}~\bibnamefont{Li}}, \bibinfo{journal}{Opt.
  Express} \textbf{\bibinfo{volume}{30}}, \bibinfo{pages}{7987}
  (\bibinfo{year}{2022}).

\bibitem[{\citenamefont{Ying et~al.}(2022)\citenamefont{Ying, Guo, Li, Gong,
  Deng, Chen, Zha, Ye, Wang, Zhu et~al.}}]{PhysRevA.105.012418}
\bibinfo{author}{\bibfnamefont{C.}~\bibnamefont{Ying}},
  \bibinfo{author}{\bibfnamefont{Q.}~\bibnamefont{Guo}},
  \bibinfo{author}{\bibfnamefont{S.}~\bibnamefont{Li}},
  \bibinfo{author}{\bibfnamefont{M.}~\bibnamefont{Gong}},
  \bibinfo{author}{\bibfnamefont{X.-H.} \bibnamefont{Deng}},
  \bibinfo{author}{\bibfnamefont{F.}~\bibnamefont{Chen}},
  \bibinfo{author}{\bibfnamefont{C.}~\bibnamefont{Zha}},
  \bibinfo{author}{\bibfnamefont{Y.}~\bibnamefont{Ye}},
  \bibinfo{author}{\bibfnamefont{C.}~\bibnamefont{Wang}},
  \bibinfo{author}{\bibfnamefont{Q.}~\bibnamefont{Zhu}}, \bibnamefont{et~al.},
  \bibinfo{journal}{Phys. Rev. A} \textbf{\bibinfo{volume}{105}},
  \bibinfo{pages}{012418} (\bibinfo{year}{2022}).

\bibitem[{\citenamefont{Dai et~al.}(2016)\citenamefont{Dai, Shi, and
  Yi}}]{PhysRevA.93.032121}
\bibinfo{author}{\bibfnamefont{C.~M.} \bibnamefont{Dai}},
  \bibinfo{author}{\bibfnamefont{Z.~C.} \bibnamefont{Shi}}, \bibnamefont{and}
  \bibinfo{author}{\bibfnamefont{X.~X.} \bibnamefont{Yi}},
  \bibinfo{journal}{Phys. Rev. A} \textbf{\bibinfo{volume}{93}},
  \bibinfo{pages}{032121} (\bibinfo{year}{2016}).

\bibitem[{\citenamefont{Harris et~al.}(1990)\citenamefont{Harris, Field, and
  Imamo\ifmmode~\breve{g}\else \u{g}\fi{}lu}}]{PhysRevLett.64.1107}
\bibinfo{author}{\bibfnamefont{S.~E.} \bibnamefont{Harris}},
  \bibinfo{author}{\bibfnamefont{J.~E.} \bibnamefont{Field}}, \bibnamefont{and}
  \bibinfo{author}{\bibfnamefont{A.}~\bibnamefont{Imamo\ifmmode~\breve{g}\else
  \u{g}\fi{}lu}}, \bibinfo{journal}{Phys. Rev. Lett.}
  \textbf{\bibinfo{volume}{64}}, \bibinfo{pages}{1107} (\bibinfo{year}{1990}).

\bibitem[{\citenamefont{Ian et~al.}(2010)\citenamefont{Ian, Liu, and
  Nori}}]{PhysRevA.81.063823}
\bibinfo{author}{\bibfnamefont{H.}~\bibnamefont{Ian}},
  \bibinfo{author}{\bibfnamefont{Y.-x.} \bibnamefont{Liu}}, \bibnamefont{and}
  \bibinfo{author}{\bibfnamefont{F.}~\bibnamefont{Nori}},
  \bibinfo{journal}{Phys. Rev. A} \textbf{\bibinfo{volume}{81}},
  \bibinfo{pages}{063823} (\bibinfo{year}{2010}).

\bibitem[{\citenamefont{Autler and Townes}(1955)}]{PhysRev.100.703}
\bibinfo{author}{\bibfnamefont{S.~H.} \bibnamefont{Autler}} \bibnamefont{and}
  \bibinfo{author}{\bibfnamefont{C.~H.} \bibnamefont{Townes}},
  \bibinfo{journal}{Phys. Rev.} \textbf{\bibinfo{volume}{100}},
  \bibinfo{pages}{703} (\bibinfo{year}{1955}).

\bibitem[{\citenamefont{Zhang et~al.}(2011)\citenamefont{Zhang, Li, Zheng,
  Wang, Chen, Li, Zhang, and Xiao}}]{2011Observation}
\bibinfo{author}{\bibfnamefont{Y.}~\bibnamefont{Zhang}},
  \bibinfo{author}{\bibfnamefont{P.}~\bibnamefont{Li}},
  \bibinfo{author}{\bibfnamefont{H.}~\bibnamefont{Zheng}},
  \bibinfo{author}{\bibfnamefont{Z.}~\bibnamefont{Wang}},
  \bibinfo{author}{\bibfnamefont{H.}~\bibnamefont{Chen}},
  \bibinfo{author}{\bibfnamefont{C.}~\bibnamefont{Li}},
  \bibinfo{author}{\bibfnamefont{R.}~\bibnamefont{Zhang}}, \bibnamefont{and}
  \bibinfo{author}{\bibfnamefont{M.}~\bibnamefont{Xiao}},
  \bibinfo{journal}{Opt. Exp.} \textbf{\bibinfo{volume}{19}},
  \bibinfo{pages}{7769} (\bibinfo{year}{2011}).

\bibitem[{\citenamefont{Abi-Salloum}(2010)}]{PhysRevA.81.053836}
\bibinfo{author}{\bibfnamefont{T.~Y.} \bibnamefont{Abi-Salloum}},
  \bibinfo{journal}{Phys. Rev. A} \textbf{\bibinfo{volume}{81}},
  \bibinfo{pages}{053836} (\bibinfo{year}{2010}).

\bibitem[{\citenamefont{Zhu et~al.}(2013)\citenamefont{Zhu, Tan, and
  Huang}}]{PhysRevA.87.043813}
\bibinfo{author}{\bibfnamefont{C.}~\bibnamefont{Zhu}},
  \bibinfo{author}{\bibfnamefont{C.}~\bibnamefont{Tan}}, \bibnamefont{and}
  \bibinfo{author}{\bibfnamefont{G.}~\bibnamefont{Huang}},
  \bibinfo{journal}{Phys. Rev. A} \textbf{\bibinfo{volume}{87}},
  \bibinfo{pages}{043813} (\bibinfo{year}{2013}).

\bibitem[{\citenamefont{Sun et~al.}(2014)\citenamefont{Sun, Liu, Ian, You,
  Il'ichev, and Nori}}]{PhysRevA.89.063822}
\bibinfo{author}{\bibfnamefont{H.-C.} \bibnamefont{Sun}},
  \bibinfo{author}{\bibfnamefont{Y.-x.} \bibnamefont{Liu}},
  \bibinfo{author}{\bibfnamefont{H.}~\bibnamefont{Ian}},
  \bibinfo{author}{\bibfnamefont{J.~Q.} \bibnamefont{You}},
  \bibinfo{author}{\bibfnamefont{E.}~\bibnamefont{Il'ichev}}, \bibnamefont{and}
  \bibinfo{author}{\bibfnamefont{F.}~\bibnamefont{Nori}},
  \bibinfo{journal}{Phys. Rev. A} \textbf{\bibinfo{volume}{89}},
  \bibinfo{pages}{063822} (\bibinfo{year}{2014}).

\bibitem[{\citenamefont{Zhao et~al.}(2022)\citenamefont{Zhao, Zhang, and
  Wang}}]{2022Phase}
\bibinfo{author}{\bibfnamefont{W.}~\bibnamefont{Zhao}},
  \bibinfo{author}{\bibfnamefont{Y.}~\bibnamefont{Zhang}}, \bibnamefont{and}
  \bibinfo{author}{\bibfnamefont{Z.}~\bibnamefont{Wang}},
  \bibinfo{journal}{Frontiers of Physics} \textbf{\bibinfo{volume}{17}},
  \bibinfo{pages}{42506} (\bibinfo{year}{2022}).

\bibitem[{\citenamefont{Wanare}(2006)}]{PhysRevLett.96.183601}
\bibinfo{author}{\bibfnamefont{H.}~\bibnamefont{Wanare}},
  \bibinfo{journal}{Phys. Rev. Lett.} \textbf{\bibinfo{volume}{96}},
  \bibinfo{pages}{183601} (\bibinfo{year}{2006}).

\bibitem[{\citenamefont{Holthaus and Just}(1994)}]{PhysRevA.49.1950}
\bibinfo{author}{\bibfnamefont{M.}~\bibnamefont{Holthaus}} \bibnamefont{and}
  \bibinfo{author}{\bibfnamefont{B.}~\bibnamefont{Just}},
  \bibinfo{journal}{Phys. Rev. A} \textbf{\bibinfo{volume}{49}},
  \bibinfo{pages}{1950} (\bibinfo{year}{1994}).

\bibitem[{\citenamefont{Picque and Hansch}(2019)}]{Picqu2019Frequency}
\bibinfo{author}{\bibfnamefont{N.}~\bibnamefont{Picque}} \bibnamefont{and}
  \bibinfo{author}{\bibfnamefont{T.~W.} \bibnamefont{Hansch}},
  \bibinfo{journal}{Nat. Photonics} \textbf{\bibinfo{volume}{13}},
  \bibinfo{pages}{146} (\bibinfo{year}{2019}).

\bibitem[{\citenamefont{Tan et~al.}(2014)\citenamefont{Tan, Zhang, Zhang, Yu,
  Han, and Zhu}}]{PhysRevLett.112.027001}
\bibinfo{author}{\bibfnamefont{X.}~\bibnamefont{Tan}},
  \bibinfo{author}{\bibfnamefont{D.-W.} \bibnamefont{Zhang}},
  \bibinfo{author}{\bibfnamefont{Z.}~\bibnamefont{Zhang}},
  \bibinfo{author}{\bibfnamefont{Y.}~\bibnamefont{Yu}},
  \bibinfo{author}{\bibfnamefont{S.}~\bibnamefont{Han}}, \bibnamefont{and}
  \bibinfo{author}{\bibfnamefont{S.-L.} \bibnamefont{Zhu}},
  \bibinfo{journal}{Phys. Rev. Lett.} \textbf{\bibinfo{volume}{112}},
  \bibinfo{pages}{027001} (\bibinfo{year}{2014}).

\bibitem[{\citenamefont{Anisimov et~al.}(2011)\citenamefont{Anisimov, Dowling,
  and Sanders}}]{PhysRevLett.107.163604}
\bibinfo{author}{\bibfnamefont{P.~M.} \bibnamefont{Anisimov}},
  \bibinfo{author}{\bibfnamefont{J.~P.} \bibnamefont{Dowling}},
  \bibnamefont{and} \bibinfo{author}{\bibfnamefont{B.~C.}
  \bibnamefont{Sanders}}, \bibinfo{journal}{Phys. Rev. Lett.}
  \textbf{\bibinfo{volume}{107}}, \bibinfo{pages}{163604}
  (\bibinfo{year}{2011}).

\bibitem[{\citenamefont{Schnell et~al.}(2021)\citenamefont{Schnell, Denisov,
  and Eckardt}}]{PhysRevB.104.165414}
\bibinfo{author}{\bibfnamefont{A.}~\bibnamefont{Schnell}},
  \bibinfo{author}{\bibfnamefont{S.}~\bibnamefont{Denisov}}, \bibnamefont{and}
  \bibinfo{author}{\bibfnamefont{A.}~\bibnamefont{Eckardt}},
  \bibinfo{journal}{Phys. Rev. B} \textbf{\bibinfo{volume}{104}},
  \bibinfo{pages}{165414} (\bibinfo{year}{2021}).

\bibitem[{\citenamefont{Schnell et~al.}(2020)\citenamefont{Schnell, Eckardt,
  and Denisov}}]{PhysRevB.101.100301}
\bibinfo{author}{\bibfnamefont{A.}~\bibnamefont{Schnell}},
  \bibinfo{author}{\bibfnamefont{A.}~\bibnamefont{Eckardt}}, \bibnamefont{and}
  \bibinfo{author}{\bibfnamefont{S.}~\bibnamefont{Denisov}},
  \bibinfo{journal}{Phys. Rev. B} \textbf{\bibinfo{volume}{101}},
  \bibinfo{pages}{100301(R)} (\bibinfo{year}{2020}).

\bibitem[{\citenamefont{Liu et~al.}(2016)\citenamefont{Liu, Li, Luo, Zhao,
  Xiong, Zhang, Chen, Liu, Chen, Nori et~al.}}]{PhysRevA.93.053838}
\bibinfo{author}{\bibfnamefont{Q.-C.} \bibnamefont{Liu}},
  \bibinfo{author}{\bibfnamefont{T.-F.} \bibnamefont{Li}},
  \bibinfo{author}{\bibfnamefont{X.-Q.} \bibnamefont{Luo}},
  \bibinfo{author}{\bibfnamefont{H.}~\bibnamefont{Zhao}},
  \bibinfo{author}{\bibfnamefont{W.}~\bibnamefont{Xiong}},
  \bibinfo{author}{\bibfnamefont{Y.-S.} \bibnamefont{Zhang}},
  \bibinfo{author}{\bibfnamefont{Z.}~\bibnamefont{Chen}},
  \bibinfo{author}{\bibfnamefont{J.~S.} \bibnamefont{Liu}},
  \bibinfo{author}{\bibfnamefont{W.}~\bibnamefont{Chen}},
  \bibinfo{author}{\bibfnamefont{F.}~\bibnamefont{Nori}}, \bibnamefont{et~al.},
  \bibinfo{journal}{Phys. Rev. A} \textbf{\bibinfo{volume}{93}},
  \bibinfo{pages}{053838} (\bibinfo{year}{2016}).

\bibitem[{\citenamefont{Grossmann et~al.}(1991)\citenamefont{Grossmann,
  Dittrich, Jung, and H\"anggi}}]{PhysRevLett.67.516}
\bibinfo{author}{\bibfnamefont{F.}~\bibnamefont{Grossmann}},
  \bibinfo{author}{\bibfnamefont{T.}~\bibnamefont{Dittrich}},
  \bibinfo{author}{\bibfnamefont{P.}~\bibnamefont{Jung}}, \bibnamefont{and}
  \bibinfo{author}{\bibfnamefont{P.}~\bibnamefont{H\"anggi}},
  \bibinfo{journal}{Phys. Rev. Lett.} \textbf{\bibinfo{volume}{67}},
  \bibinfo{pages}{516} (\bibinfo{year}{1991}).

\bibitem[{\citenamefont{Ashhab et~al.}(2007)\citenamefont{Ashhab, Johansson,
  Zagoskin, and Nori}}]{PhysRevA.75.063414}
\bibinfo{author}{\bibfnamefont{S.}~\bibnamefont{Ashhab}},
  \bibinfo{author}{\bibfnamefont{J.~R.} \bibnamefont{Johansson}},
  \bibinfo{author}{\bibfnamefont{A.~M.} \bibnamefont{Zagoskin}},
  \bibnamefont{and} \bibinfo{author}{\bibfnamefont{F.}~\bibnamefont{Nori}},
  \bibinfo{journal}{Phys. Rev. A} \textbf{\bibinfo{volume}{75}},
  \bibinfo{pages}{063414} (\bibinfo{year}{2007}).

\bibitem[{\citenamefont{Li et~al.}(2015)\citenamefont{Li, Luo, L\"u, Yang, and
  Wu}}]{PhysRevA.91.063804}
\bibinfo{author}{\bibfnamefont{L.}~\bibnamefont{Li}},
  \bibinfo{author}{\bibfnamefont{X.}~\bibnamefont{Luo}},
  \bibinfo{author}{\bibfnamefont{X.-Y.} \bibnamefont{L\"u}},
  \bibinfo{author}{\bibfnamefont{X.}~\bibnamefont{Yang}}, \bibnamefont{and}
  \bibinfo{author}{\bibfnamefont{Y.}~\bibnamefont{Wu}}, \bibinfo{journal}{Phys.
  Rev. A} \textbf{\bibinfo{volume}{91}}, \bibinfo{pages}{063804}
  (\bibinfo{year}{2015}).

\bibitem[{\citenamefont{Hu et~al.}(2006)\citenamefont{Hu, Xu, Li, Li, Shi, and
  Zhang}}]{2006Bichromatic}
\bibinfo{author}{\bibfnamefont{X.~M.} \bibnamefont{Hu}},
  \bibinfo{author}{\bibfnamefont{Q.}~\bibnamefont{Xu}},
  \bibinfo{author}{\bibfnamefont{J.~Y.} \bibnamefont{Li}},
  \bibinfo{author}{\bibfnamefont{X.~X.} \bibnamefont{Li}},
  \bibinfo{author}{\bibfnamefont{W.~X.} \bibnamefont{Shi}}, \bibnamefont{and}
  \bibinfo{author}{\bibfnamefont{X.}~\bibnamefont{Zhang}},
  \bibinfo{journal}{Opt. Commun.} \textbf{\bibinfo{volume}{260}},
  \bibinfo{pages}{196} (\bibinfo{year}{2006}).

\bibitem[{\citenamefont{Jakob and Kryuchkyan}(1998)}]{PhysRevA.57.1355}
\bibinfo{author}{\bibfnamefont{M.}~\bibnamefont{Jakob}} \bibnamefont{and}
  \bibinfo{author}{\bibfnamefont{G.~Y.} \bibnamefont{Kryuchkyan}},
  \bibinfo{journal}{Phys. Rev. A} \textbf{\bibinfo{volume}{57}},
  \bibinfo{pages}{1355} (\bibinfo{year}{1998}).

\bibitem[{\citenamefont{Wang et~al.}(2003)\citenamefont{Wang, Zhu, Jiang, and
  Zhan}}]{PhysRevA.68.063810}
\bibinfo{author}{\bibfnamefont{J.}~\bibnamefont{Wang}},
  \bibinfo{author}{\bibfnamefont{Y.}~\bibnamefont{Zhu}},
  \bibinfo{author}{\bibfnamefont{K.~J.} \bibnamefont{Jiang}}, \bibnamefont{and}
  \bibinfo{author}{\bibfnamefont{M.~S.} \bibnamefont{Zhan}},
  \bibinfo{journal}{Phys. Rev. A} \textbf{\bibinfo{volume}{68}},
  \bibinfo{pages}{063810} (\bibinfo{year}{2003}).

\bibitem[{\citenamefont{Hsiao et~al.}(2018)\citenamefont{Hsiao, Tsai, Chen,
  Lin, Hung, Lee, Chen, Chen, Yu, and Chen}}]{PhysRevLett.120.183602}
\bibinfo{author}{\bibfnamefont{Y.-F.} \bibnamefont{Hsiao}},
  \bibinfo{author}{\bibfnamefont{P.-J.} \bibnamefont{Tsai}},
  \bibinfo{author}{\bibfnamefont{H.-S.} \bibnamefont{Chen}},
  \bibinfo{author}{\bibfnamefont{S.-X.} \bibnamefont{Lin}},
  \bibinfo{author}{\bibfnamefont{C.-C.} \bibnamefont{Hung}},
  \bibinfo{author}{\bibfnamefont{C.-H.} \bibnamefont{Lee}},
  \bibinfo{author}{\bibfnamefont{Y.-H.} \bibnamefont{Chen}},
  \bibinfo{author}{\bibfnamefont{Y.-F.} \bibnamefont{Chen}},
  \bibinfo{author}{\bibfnamefont{I.~A.} \bibnamefont{Yu}}, \bibnamefont{and}
  \bibinfo{author}{\bibfnamefont{Y.-C.} \bibnamefont{Chen}},
  \bibinfo{journal}{Phys. Rev. Lett.} \textbf{\bibinfo{volume}{120}},
  \bibinfo{pages}{183602} (\bibinfo{year}{2018}).

\bibitem[{\citenamefont{Yi et~al.}(2013)\citenamefont{Yi, Gu, and
  Li}}]{YiGround}
\bibinfo{author}{\bibfnamefont{Z.}~\bibnamefont{Yi}},
  \bibinfo{author}{\bibfnamefont{W.-j.} \bibnamefont{Gu}}, \bibnamefont{and}
  \bibinfo{author}{\bibfnamefont{G.-x.} \bibnamefont{Li}},
  \bibinfo{journal}{Opt. Exp.} \textbf{\bibinfo{volume}{21}},
  \bibinfo{pages}{3445} (\bibinfo{year}{2013}).

\bibitem[{\citenamefont{Peng et~al.}(2014)\citenamefont{Peng, Özdemir SK,
  Chen, Nori, and Yang}}]{Peng2014EIT}
\bibinfo{author}{\bibfnamefont{B.}~\bibnamefont{Peng}},
  \bibinfo{author}{\bibnamefont{Özdemir SK}},
  \bibinfo{author}{\bibfnamefont{W.}~\bibnamefont{Chen}},
  \bibinfo{author}{\bibfnamefont{F.}~\bibnamefont{Nori}}, \bibnamefont{and}
  \bibinfo{author}{\bibfnamefont{L.}~\bibnamefont{Yang}},
  \bibinfo{journal}{Nat. Commun.} \textbf{\bibinfo{volume}{5}},
  \bibinfo{pages}{5082} (\bibinfo{year}{2014}).

\bibitem[{\citenamefont{Ling-Yan et~al.}(2015)\citenamefont{Ling-Yan, He,
  Tie-Jun, Wang, Yong-Pan, Gao, Cong, Cao, and Chuan}}]{Ling2015Discerning}
\bibinfo{author}{\bibnamefont{Ling-Yan}}, \bibinfo{author}{\bibnamefont{He}},
  \bibinfo{author}{\bibnamefont{Tie-Jun}},
  \bibinfo{author}{\bibnamefont{Wang}},
  \bibinfo{author}{\bibnamefont{Yong-Pan}},
  \bibinfo{author}{\bibnamefont{Gao}}, \bibinfo{author}{\bibnamefont{Cong}},
  \bibinfo{author}{\bibnamefont{Cao}}, \bibnamefont{and}
  \bibinfo{author}{\bibnamefont{Chuan}}, \bibinfo{journal}{Opt. Exp.}
  \textbf{\bibinfo{volume}{23}}, \bibinfo{pages}{23817} (\bibinfo{year}{2015}).

\bibitem[{\citenamefont{Fleischhauer et~al.}(2005)\citenamefont{Fleischhauer,
  Imamoglu, and Marangos}}]{RevModPhys.77.633}
\bibinfo{author}{\bibfnamefont{M.}~\bibnamefont{Fleischhauer}},
  \bibinfo{author}{\bibfnamefont{A.}~\bibnamefont{Imamoglu}}, \bibnamefont{and}
  \bibinfo{author}{\bibfnamefont{J.~P.} \bibnamefont{Marangos}},
  \bibinfo{journal}{Rev. Mod. Phys.} \textbf{\bibinfo{volume}{77}},
  \bibinfo{pages}{633} (\bibinfo{year}{2005}).

\bibitem[{\citenamefont{Davuluri and Zhu}(2016)}]{2016Interference}
\bibinfo{author}{\bibfnamefont{S.}~\bibnamefont{Davuluri}} \bibnamefont{and}
  \bibinfo{author}{\bibfnamefont{S.}~\bibnamefont{Zhu}},
  \bibinfo{journal}{Phys. Scr.} \textbf{\bibinfo{volume}{91}},
  \bibinfo{pages}{013008} (\bibinfo{year}{2016}).

\bibitem[{\citenamefont{Liu et~al.}(2020)\citenamefont{Liu, Wu, Zhang, He, and
  Zhang}}]{Liu:20}
\bibinfo{author}{\bibfnamefont{J.}~\bibnamefont{Liu}},
  \bibinfo{author}{\bibfnamefont{J.}~\bibnamefont{Wu}},
  \bibinfo{author}{\bibfnamefont{Y.}~\bibnamefont{Zhang}},
  \bibinfo{author}{\bibfnamefont{Y.}~\bibnamefont{He}}, \bibnamefont{and}
  \bibinfo{author}{\bibfnamefont{J.}~\bibnamefont{Zhang}}, \bibinfo{journal}{J.
  Opt. Soc. Am. B} \textbf{\bibinfo{volume}{37}}, \bibinfo{pages}{49}
  (\bibinfo{year}{2020}).

\bibitem[{\citenamefont{Itano et~al.}(1990)\citenamefont{Itano, Heinzen,
  Bollinger, and Wineland}}]{PhysRevA.41.2295}
\bibinfo{author}{\bibfnamefont{W.~M.} \bibnamefont{Itano}},
  \bibinfo{author}{\bibfnamefont{D.~J.} \bibnamefont{Heinzen}},
  \bibinfo{author}{\bibfnamefont{J.~J.} \bibnamefont{Bollinger}},
  \bibnamefont{and} \bibinfo{author}{\bibfnamefont{D.~J.}
  \bibnamefont{Wineland}}, \bibinfo{journal}{Phys. Rev. A}
  \textbf{\bibinfo{volume}{41}}, \bibinfo{pages}{2295} (\bibinfo{year}{1990}).

\bibitem[{\citenamefont{Rubio-Abadal et~al.}(2020)\citenamefont{Rubio-Abadal,
  Ippoliti, Hollerith, Wei, Rui, Sondhi, Khemani, Gross, and
  Bloch}}]{PhysRevX.10.021044}
\bibinfo{author}{\bibfnamefont{A.}~\bibnamefont{Rubio-Abadal}},
  \bibinfo{author}{\bibfnamefont{M.}~\bibnamefont{Ippoliti}},
  \bibinfo{author}{\bibfnamefont{S.}~\bibnamefont{Hollerith}},
  \bibinfo{author}{\bibfnamefont{D.}~\bibnamefont{Wei}},
  \bibinfo{author}{\bibfnamefont{J.}~\bibnamefont{Rui}},
  \bibinfo{author}{\bibfnamefont{S.~L.} \bibnamefont{Sondhi}},
  \bibinfo{author}{\bibfnamefont{V.}~\bibnamefont{Khemani}},
  \bibinfo{author}{\bibfnamefont{C.}~\bibnamefont{Gross}}, \bibnamefont{and}
  \bibinfo{author}{\bibfnamefont{I.}~\bibnamefont{Bloch}},
  \bibinfo{journal}{Phys. Rev. X} \textbf{\bibinfo{volume}{10}},
  \bibinfo{pages}{021044} (\bibinfo{year}{2020}).

\bibitem[{\citenamefont{Lukin et~al.}(1997)\citenamefont{Lukin, Fleischhauer,
  Zibrov, Robinson, Velichansky, Hollberg, and Scully}}]{PhysRevLett.79.2959}
\bibinfo{author}{\bibfnamefont{M.~D.} \bibnamefont{Lukin}},
  \bibinfo{author}{\bibfnamefont{M.}~\bibnamefont{Fleischhauer}},
  \bibinfo{author}{\bibfnamefont{A.~S.} \bibnamefont{Zibrov}},
  \bibinfo{author}{\bibfnamefont{H.~G.} \bibnamefont{Robinson}},
  \bibinfo{author}{\bibfnamefont{V.~L.} \bibnamefont{Velichansky}},
  \bibinfo{author}{\bibfnamefont{L.}~\bibnamefont{Hollberg}}, \bibnamefont{and}
  \bibinfo{author}{\bibfnamefont{M.~O.} \bibnamefont{Scully}},
  \bibinfo{journal}{Phys. Rev. Lett.} \textbf{\bibinfo{volume}{79}},
  \bibinfo{pages}{2959} (\bibinfo{year}{1997}).

\bibitem[{\citenamefont{Novikova et~al.}(2012)\citenamefont{Novikova,
  Walsworth, and Xiao}}]{NovikovaElectromagnetically}
\bibinfo{author}{\bibfnamefont{I.}~\bibnamefont{Novikova}},
  \bibinfo{author}{\bibfnamefont{R.}~\bibnamefont{Walsworth}},
  \bibnamefont{and} \bibinfo{author}{\bibfnamefont{Y.}~\bibnamefont{Xiao}},
  \bibinfo{journal}{Laser Photonics Rev.} \textbf{\bibinfo{volume}{6}},
  \bibinfo{pages}{333} (\bibinfo{year}{2012}).

\bibitem[{\citenamefont{You and Nori}(2011)}]{You2011Atomic}
\bibinfo{author}{\bibfnamefont{J.~Q.} \bibnamefont{You}} \bibnamefont{and}
  \bibinfo{author}{\bibfnamefont{F.}~\bibnamefont{Nori}},
  \bibinfo{journal}{Nature} \textbf{\bibinfo{volume}{474}},
  \bibinfo{pages}{589} (\bibinfo{year}{2011}).

\bibitem[{\citenamefont{Xu et~al.}(2007)\citenamefont{Xu, Sun, Berman, Steel,
  Bracker, Gammon, and Sham}}]{XuCoherent}
\bibinfo{author}{\bibfnamefont{X.}~\bibnamefont{Xu}},
  \bibinfo{author}{\bibfnamefont{B.}~\bibnamefont{Sun}},
  \bibinfo{author}{\bibfnamefont{P.~R.} \bibnamefont{Berman}},
  \bibinfo{author}{\bibfnamefont{D.~G.} \bibnamefont{Steel}},
  \bibinfo{author}{\bibfnamefont{A.~S.} \bibnamefont{Bracker}},
  \bibinfo{author}{\bibfnamefont{D.}~\bibnamefont{Gammon}}, \bibnamefont{and}
  \bibinfo{author}{\bibfnamefont{L.~J.} \bibnamefont{Sham}},
  \bibinfo{journal}{science} \textbf{\bibinfo{volume}{317}},
  \bibinfo{pages}{929} (\bibinfo{year}{2007}).

\bibitem[{\citenamefont{Liu et~al.}(2009)\citenamefont{Liu, Langguth, Weiss,
  K\"astel, Fleischhauer, Pfau, and Giessen}}]{LiuPlasmonic}
\bibinfo{author}{\bibfnamefont{N.}~\bibnamefont{Liu}},
  \bibinfo{author}{\bibfnamefont{L.}~\bibnamefont{Langguth}},
  \bibinfo{author}{\bibfnamefont{T.}~\bibnamefont{Weiss}},
  \bibinfo{author}{\bibfnamefont{J.}~\bibnamefont{K\"astel}},
  \bibinfo{author}{\bibfnamefont{M.}~\bibnamefont{Fleischhauer}},
  \bibinfo{author}{\bibfnamefont{T.}~\bibnamefont{Pfau}}, \bibnamefont{and}
  \bibinfo{author}{\bibfnamefont{H.}~\bibnamefont{Giessen}},
  \bibinfo{journal}{Nat. Mater.} \textbf{\bibinfo{volume}{8}},
  \bibinfo{pages}{758} (\bibinfo{year}{2009}).

\bibitem[{\citenamefont{Safavi-Naeini et~al.}(2011)\citenamefont{Safavi-Naeini,
  Alegre, Chan, Eichenfield, Winger, Lin, Hill, Chang, and
  Painter}}]{safavi2011}
\bibinfo{author}{\bibfnamefont{A.~H.} \bibnamefont{Safavi-Naeini}},
  \bibinfo{author}{\bibfnamefont{T.~P.~M.} \bibnamefont{Alegre}},
  \bibinfo{author}{\bibfnamefont{J.}~\bibnamefont{Chan}},
  \bibinfo{author}{\bibfnamefont{M.}~\bibnamefont{Eichenfield}},
  \bibinfo{author}{\bibfnamefont{M.}~\bibnamefont{Winger}},
  \bibinfo{author}{\bibfnamefont{Q.}~\bibnamefont{Lin}},
  \bibinfo{author}{\bibfnamefont{J.~T.} \bibnamefont{Hill}},
  \bibinfo{author}{\bibfnamefont{D.~E.} \bibnamefont{Chang}}, \bibnamefont{and}
  \bibinfo{author}{\bibfnamefont{O.}~\bibnamefont{Painter}}
  (\bibinfo{year}{2011}), vol. \bibinfo{volume}{472}, pp.
  \bibinfo{pages}{69--73}.

\bibitem[{\citenamefont{Jiang et~al.}(2013)\citenamefont{Jiang, Liu, Cui, Li,
  Chen, and Chen}}]{JiangElectromagnetically}
\bibinfo{author}{\bibfnamefont{C.}~\bibnamefont{Jiang}},
  \bibinfo{author}{\bibfnamefont{H.}~\bibnamefont{Liu}},
  \bibinfo{author}{\bibfnamefont{Y.}~\bibnamefont{Cui}},
  \bibinfo{author}{\bibfnamefont{X.}~\bibnamefont{Li}},
  \bibinfo{author}{\bibfnamefont{G.}~\bibnamefont{Chen}}, \bibnamefont{and}
  \bibinfo{author}{\bibfnamefont{B.}~\bibnamefont{Chen}},
  \bibinfo{journal}{Opt. Express} \textbf{\bibinfo{volume}{21}},
  \bibinfo{pages}{12165} (\bibinfo{year}{2013}).

\bibitem[{\citenamefont{Wang et~al.}(2022)\citenamefont{Wang, Li, Xu, Li, Wang,
  Yang, Mi, Liang, Su, Yang et~al.}}]{WOS:000742352300001}
\bibinfo{author}{\bibfnamefont{C.}~\bibnamefont{Wang}},
  \bibinfo{author}{\bibfnamefont{X.}~\bibnamefont{Li}},
  \bibinfo{author}{\bibfnamefont{H.}~\bibnamefont{Xu}},
  \bibinfo{author}{\bibfnamefont{Z.}~\bibnamefont{Li}},
  \bibinfo{author}{\bibfnamefont{J.}~\bibnamefont{Wang}},
  \bibinfo{author}{\bibfnamefont{Z.}~\bibnamefont{Yang}},
  \bibinfo{author}{\bibfnamefont{Z.}~\bibnamefont{Mi}},
  \bibinfo{author}{\bibfnamefont{X.}~\bibnamefont{Liang}},
  \bibinfo{author}{\bibfnamefont{T.}~\bibnamefont{Su}},
  \bibinfo{author}{\bibfnamefont{C.}~\bibnamefont{Yang}}, \bibnamefont{et~al.},
  \bibinfo{journal}{npj Quantum Inf.} \textbf{\bibinfo{volume}{8}}
  (\bibinfo{year}{2022}).

\bibitem[{\citenamefont{Paik et~al.}(2011)\citenamefont{Paik, Schuster, Bishop,
  Kirchmair, Catelani, Sears, Johnson, Reagor, Frunzio, Glazman
  et~al.}}]{PhysRevLett.107.240501}
\bibinfo{author}{\bibfnamefont{H.}~\bibnamefont{Paik}},
  \bibinfo{author}{\bibfnamefont{D.~I.} \bibnamefont{Schuster}},
  \bibinfo{author}{\bibfnamefont{L.~S.} \bibnamefont{Bishop}},
  \bibinfo{author}{\bibfnamefont{G.}~\bibnamefont{Kirchmair}},
  \bibinfo{author}{\bibfnamefont{G.}~\bibnamefont{Catelani}},
  \bibinfo{author}{\bibfnamefont{A.~P.} \bibnamefont{Sears}},
  \bibinfo{author}{\bibfnamefont{B.~R.} \bibnamefont{Johnson}},
  \bibinfo{author}{\bibfnamefont{M.~J.} \bibnamefont{Reagor}},
  \bibinfo{author}{\bibfnamefont{L.}~\bibnamefont{Frunzio}},
  \bibinfo{author}{\bibfnamefont{L.~I.} \bibnamefont{Glazman}},
  \bibnamefont{et~al.}, \bibinfo{journal}{Phys. Rev. Lett.}
  \textbf{\bibinfo{volume}{107}}, \bibinfo{pages}{240501}
  (\bibinfo{year}{2011}).

\bibitem[{\citenamefont{Rigetti et~al.}(2012)\citenamefont{Rigetti, Gambetta,
  Poletto, Plourde, Chow, C\'orcoles, Smolin, Merkel, Rozen, Keefe
  et~al.}}]{PhysRevB.86.100506}
\bibinfo{author}{\bibfnamefont{C.}~\bibnamefont{Rigetti}},
  \bibinfo{author}{\bibfnamefont{J.~M.} \bibnamefont{Gambetta}},
  \bibinfo{author}{\bibfnamefont{S.}~\bibnamefont{Poletto}},
  \bibinfo{author}{\bibfnamefont{B.~L.~T.} \bibnamefont{Plourde}},
  \bibinfo{author}{\bibfnamefont{J.~M.} \bibnamefont{Chow}},
  \bibinfo{author}{\bibfnamefont{A.~D.} \bibnamefont{C\'orcoles}},
  \bibinfo{author}{\bibfnamefont{J.~A.} \bibnamefont{Smolin}},
  \bibinfo{author}{\bibfnamefont{S.~T.} \bibnamefont{Merkel}},
  \bibinfo{author}{\bibfnamefont{J.~R.} \bibnamefont{Rozen}},
  \bibinfo{author}{\bibfnamefont{G.~A.} \bibnamefont{Keefe}},
  \bibnamefont{et~al.}, \bibinfo{journal}{Phys. Rev. B}
  \textbf{\bibinfo{volume}{86}}, \bibinfo{pages}{100506(R)}
  (\bibinfo{year}{2012}).

\bibitem[{\citenamefont{Ficek et~al.}(2000)\citenamefont{Ficek, Seke, Soldatov,
  and Adam}}]{Ficek2000Saturation}
\bibinfo{author}{\bibfnamefont{Z.}~\bibnamefont{Ficek}},
  \bibinfo{author}{\bibfnamefont{J.}~\bibnamefont{Seke}},
  \bibinfo{author}{\bibfnamefont{A.}~\bibnamefont{Soldatov}}, \bibnamefont{and}
  \bibinfo{author}{\bibfnamefont{G.}~\bibnamefont{Adam}}, \bibinfo{journal}{J.
  Opt. B} \textbf{\bibinfo{volume}{2}}, \bibinfo{pages}{780}
  (\bibinfo{year}{2000}).

\end{thebibliography}

\end{document}